\documentclass{article}
\usepackage{arxiv}

\def\covariate{\bm{z}}

\def\hMSE{\widehat{\mathrm{MSE}}}
\def\hCI{\widehat{\mathrm{CI}}}
\def\hVAR{\widehat{\mathrm{Var}}}
\def\Bernoulli{\mathrm{Bernoulli}}
\def\M{\mathcal{M}}
\def\hVR{\widehat{\mathrm{VR}}}

\def\LIFT{\mathrm{LIFT}}
\def\ATE{\mathrm{ATE}}

\newcommand{\LLIFT}[1]{\mathrm{LIFT}(#1)}

\def\unATE{\widehat{\ATE}_{\mathrm{DIM}}}

\def\intATE{\widehat{\ATE}_{\mathrm{LR}}}

\newcommand{\fittedparam}[1]{\hat{\bm{\theta}}_{#1}}

\def\LR{\mathrm{LR}}
\def\herr{\widehat{\mathrm{err}}}

\def\DD{\mathcal{D}}

\def\hatATE{\widehat{\mathrm{ATE}}}

\def\RE{\mathbb{R}}

\def\MM{\mathcal{M}}

\newcommand{\NN}{\mathbb{N}}

\newcommand{\II}{\mathcal{I}}
\newcommand{\hatCI}[1]{\widehat{\mathrm{CI}}_{#1}}

\newcommand{\ind}{\mathbbm{1}}
\newcommand{\ES}{\mathcal{S}}

\newcommand{\hmediandist}{\widehat{\mathrm{Mediandist}}}
\newcommand{\hexcessfrac}{\widehat{\mathrm{Excessfrac}}}
\newcommand{\hcoverage}{\widehat{\mathrm{Coverage}}}

\newcommand{\DIM}{\mathrm{DIM}}

\newcommand{\hmetric}{\widehat{\mathrm{Metric}}}

\usepackage[utf8]{inputenc}
\usepackage{hyperref}
\usepackage{natbib}
\usepackage{amsmath, amssymb}
\usepackage{graphicx}
\usepackage{algpseudocode}
\usepackage{algorithm}
\usepackage{cleveref}
\usepackage{bm}
\usepackage{bbm}

\setkeys{Gin}{width=.9\textwidth}
\begin{document}

\title{Leveraging covariate adjustments at scale in online A/B testing}

\author{\begin{tabular}{ccc} Lorenzo Masoero & Doug Hains & James McQueen  \\ \texttt{masoerl@amazon.com} & \texttt{dhains@amazon.com}  & \texttt{jmcq@amazon.com}\\ & & \\ & Amazon.com & \end{tabular}}

\maketitle

\begin{abstract}
Companies offering web services routinely run randomized online experiments to estimate the ``causal impact'' associated with the adoption of new features and policies on key performance metrics of interest. These experiments are used to estimate a variety of effects: the increase in click rate due to the repositioning of a banner, the impact on subscription rate as a consequence of a discount or special offer, etc. 
In these settings, even effects whose sizes are very small can have large downstream impacts. The simple difference in means estimator \citep{splawa1990application} is still the standard estimator of choice for many online A/B testing platforms due to its simplicity. This method, however, can fail to detect small effects, even when the experiment contains thousands or millions of observational units. As a byproduct of these experiments, however, large amounts of additional data (covariates) are collected. In this paper, we discuss benefits, costs and risks of allowing experimenters to leverage more complicated estimators that make use of covariates when estimating causal effects of interest. We adapt a recently proposed general-purpose algorithm for the estimation of causal effects with covariates to the setting of online A/B testing. Through this paradigm, we implement several covariate-adjusted causal estimators. We thoroughly evaluate their performance at scale, highlighting benefits and shortcomings of different methods. We show on real experiments how ``covariate-adjusted'' estimators can (i) lead to more precise quantification of the causal effects of interest and (ii) fix issues related to imbalance across treatment arms --- a practical concern often overlooked in the literature. In turn, (iii) these more precise estimates can reduce experimentation time, cutting cost and helping to streamline decision-making processes, allowing for faster adoption of beneficial interventions.
\end{abstract}

\section{Introduction} \label{sec:intro}

The continued growth and product improvement for online companies relies on efficiently finding new opportunities and accurately measuring the impact of decisions on customers. To estimate the causal impact of a change to a product or feature, online companies heavily rely on A/B tests (randomized controlled trials/online randomized experiments). Under minimal assumption, A/B tests are indeed guaranteed to produce unbiased estimates of the causal impact of the interventions that are being tested \citep{splawa1990application}. 

In what follows, we will assume that the experimental units of interest are ``customers''. In practice, different experiments might be tracking different units (sellers, streamers, shopping missions, etc.). A/B tests work by randomly assigning some customers (usually half, the “treatment group”) to see the new experience, while the other customers (the “control group”) see the old, \emph{status quo} experience. Over a fixed experimental period, different relevant metrics of interest of these customers are measured and recorded. 
A simple way for experimenters to quantify how the change in the experience impacts customers with an A/B test is to compare the average value of a metric of interest or ``key performance indicator'' [KPI] across customers in the treatment group to the average value of the same metric in the control group. This ``difference in means'' [DIM] approach produces a simple estimate of the average causal effect of the change in the experience on the KPI in question.

The simplicity and low marginal cost of running A/B tests with millions of customers has led them to be ubiquitous in the industry. A/B testing is used to evaluate front-end and back-end changes to search engines (Google, Bing, Yandex), online retailers (Amazon, eBay, Etsy), streaming media services (Netflix, Twitch, YouTube), social networks (Facebook, LinkedIn, Twitter), travel services (Lyft, Uber, Airbnb, Booking.com), etc. See \citet{gupta2019top} for a thorough discussion of the role and use of A/B tests in the industry.  

Due to the opportunity cost of experimentation time, small treatment effect sizes, and large heterogeneity amongst customers, the difference in means approach can often fail to detect effects of the intervention when they are present, even when the experiment contains thousands or millions of customers. However, large amounts of data often unrelated with the A/B test (the ``covariates'') are collected before and throughout the experiment about the experimental units. 
This abundance of data gives experimenters the potential to adopt more complex ``covariate-adjusted'' methods to form their estimates. In particular, any feature that is independent of the intervention (such as any measures taken prior to the experiment) can be leveraged to estimate the causal effect of interest. Resulting covariate-adjusted estimators can lead to improved, less variable, estimates of the ``causal effects'' of interest. 
It has been observed empirically that simple covariate adjusted approaches, such as the popular ``CUPED'' \citep{deng2013improving}, can lead to significant variance reduction relative to the difference in means estimator. Furthermore, covariate adjustment has been used defensively to guard against an unlucky randomization, where the intervention may appear artificially better or worse due to luck (see \citet{tukey1991use}).  

The literature related to covariate adjustment methods is continuously growing (see, e.g. \citet{guo2021machine,jin2021towards} for recent contributions). In this paper, we explore benefits and costs of expanding the toolkit of experimenters by using larger sets of covariates and more complex estimators for the estimation of causal effects in the context of online A/B tests. We show in our experiments that covariate-adjusted methods can lead to non-trivial gains in terms of estimation accuracy and variance reduction. 

The rest of this paper is organized as follows: we introduce notation for the problem of interest in \Cref{sec:potential_outcomes}. Next, we describe in \Cref{sec:gobes}  the class of Generalized Oaxaca-Blinder Estimators [GOBEs] --- a flexible, general purpose method to produce estimators of the causal effects leveraging any arbitrary number of additional covariates. We discuss the potential use of these estimators for experimentation in \Cref{sec:complexity}, and present experimental results in \Cref{sec:experiments}. We conclude with a discussion and next steps in \Cref{sec:discussion}. 

\section{Potential outcomes, randomized experiments and causal effects}  \label{sec:potential_outcomes}

The field of causal inference is a collection of  theoretically sound tools, methodologies and procedures which can help practitioners answer questions about the impact of interventions they may want to implement. 
While making rigorous causal claims about interventions is appealing and desirable, this ability comes at the cost of collecting data through carefully designed experiments. In order for causal claims to be valid, experimenters have to make sure that the data is collected in such a way that no bias or flaw is introduced in the analysis. 
The standard approach to ensure that the data we are collecting will allow us to formulate causal claims, is to perform a \emph{randomized controlled trial} (RCT), or A/B test. In its simplest form,  an A/B test is implemented by exposing each experimental unit to either the control ($A$) or treatment ($B$) experience \emph{at random}. Randomization is the key technical device which allows experimenters to draw causal conclusions from the experiment. 

To make our discussion precise, we here adopt the causal model of potential outcomes \citep{splawa1990application,rubin1977assignment}. In a nutshell, we assume that in an experiment in which we observe $N$ units, every individual unit $n \in [N]:= \{1,\ldots,N\}$ is exposed to one of $T\ge 2$ different ``treatments'' or ``policies''. For example, the $N$ units might be different customers in the experiment. For each \emph{potential} allocation of unit $n\in [N]$ to policy $t$, we assume that there exists a ``potential'' outcome $Y_n(t)$. That is, each unit in the experiment is associated with a (latent) vector of potential outcomes, $[Y_n(0), \ldots, Y_n(T-1)]^\top$, of which only one coordinate is observed in an experiment. These outcomes are assumed to be fixed conditionally on the assignment. For simplicity in what follows we will consider the case of $T=2$ alternative treatments, that we will simply call ``control'' ($t=0$) or ``treatment'' ($t=1$), even though our discussion naturally extends to any $T>2$. Given these definitions, we can formally define what we mean by a  ``causal effect''. The most important effect (or estimand) of interest, and the one we will focus on, is the average treatment effect [$\ATE$]. The $\ATE$ is the (average) \emph{causal effect} in the population of exposing a unit to the treatment ($t=1$) instead of the alternative control ($t=0$). Often, for decision-making, we think of the treatment as an alternative policy to a standard baseline (potentially more expensive or riskier). The $\ATE$ quantifies the impact on the outcome of interest of adopting this alternative strategy. Formally, 
\begin{equation*}
	\bar{Y}_{k}:=\sum_{n=1}^N \frac{Y_n(k)}{N},
	\;\text{and}\; 
    \ATE:=\bar{Y}_1 - \bar{Y}_0 = \sum_{n=1}^N \frac{\left[ Y_n(1) - Y_n(0) \right]}{N}. 
\end{equation*}
The $\ATE$ can not be directly computed or observed in practice, because units are either exposed to treatment or control, but never to both. RCTs  or A/B tests are used to estimate the $\ATE$. The fundamental mechanism underlying an A/B test is its random assignment mechanism (or triggering logic), which determines the experience to which each unit will be exposed. In the simplest case, each unit $n$ is  endowed with a binary random variable with mean $\pi \in (0,1)$:
\begin{equation}
	J_n \sim \Bernoulli(\pi). \label{eq:triggering}
\end{equation}
If $J_n=0$, then unit $n$ is exposed to the control. Otherwise,  if $J_n=1$, the treatment experience is rendered. The ATE is estimated by comparing the observed outcomes for the units in control and treatment. Denote with $\II_t:=\{n \in [N]\;:\; J_n=t\}$ for the units in group $t \in \{0,1\}$. The ``difference-in-means'' [DIM] estimator is simply the difference between the average outcome in each treatment group:
\begin{equation}
    \unATE := \left[\frac{1}{|\II_1|} \sum_{n \in \II_1} Y_n(1) \right] - \left[\frac{1}{|\II_0|} \sum_{n\in\II_0} Y_n(0) \right].   \label{eq:unATE}
\end{equation}
The theoretical properties of this extremely simple estimator are well understood \citep{splawa1990application}: it is unbiased, and under mild conditions it obeys a central limit theorem in large samples. See \citet{li2017general} and the references therein for a thorough overview and discussion.


\section{Leveraging covariates: Generalized Oaxaca-Blinder estimators} \label{sec:gobes}

Often, when collecting data from our experiment, we have access to additional covariates measured at the unit level, hereafter denoted as $\bm{z}_{n} := [z_{n,1},\ldots,z_{n,K}]^\top\in\RE^K$, for $n\in[N]$ and some fixed $K\in \NN$. For example, when running an experiment on the engagement of customers subscribing to a video streaming service, we might have access to previous measurements of the customer activity, the longevity of the customer's account, whether they have subscribed to for pay-per-view channels, etc. If these covariates are:
\begin{itemize}
	\item[(C1)] independent of the assignment variable $J_n$
	\item[(C2)] correlated with the outcome variable of interest $Y_n$
\end{itemize}
they can bee leveraged to form covariate-adjusted estimators. See \citet[Chapter 7]{imbens2015causal} for a detailed discussion on the validity of regression adjustments in randomized experiments.

\par In what follows, we describe a general recipe to build ``covariate-adjusted'' estimators. The key intuition underlying this approach is to view the estimation of the $\ATE$ as a ``missing data'' or ``imputation'' problem. For each treatment $t$, we can fit a regression model using the observed data within the group $\II_t$, and use the regression to impute the ``missing'' values of units assigned to the other treatment group(s) --- $n\in \II_t^C$. That is, we fit for $t\in\{0,1\}$ a regression model $Y_n \sim f_t(\bm{z}_n;\bm{\theta}_t)$ using covariates and outcomes in the corresponding treatment group, $\DD_t:=\{(Y_n, \bm{z}_n)\}_{n\in\II_t}$. Here $\bm{\theta}_t$ is a finite dimensional parameter that characterizes the regression model (e.g., the slope and intercept of a linear regression model). Given $\DD_t$, we estimate $\bm{\theta}$ by minimizing a loss function $\mathcal{L}$ computed on $\DD_t$ and parametrized by $\bm{\theta}_t$:
\begin{equation}
    \fittedparam{t} \in \arg\min_{\theta}\mathcal{L}\left\{\DD_t;\bm{\theta}\right\}. \label{eq:minimize}
\end{equation}
This gives us the imputation operator:
\begin{equation*}
    \hat{f}_t(Y_n,\covariate_n,J_n;\fittedparam{t}) = 
    \begin{cases}
        Y_n &\mbox{ if } J_n = t, \\
        f(\bm{z}_n;\fittedparam{t}) &\mbox {otherwise.}
    \end{cases} \label{eq:imputation_operator}
\end{equation*}
This approach induces the large class of ``Generalized Oaxaca-Blinder Estimators'' [GOBEs, \citet{guo2021generalized}] of the type
\begin{equation}
    \hatATE_{\MM} = \frac{1}{N}\sum_{n=1}^N \left\{ \hat{Y}_n(1) - \hat{Y}_n(0) \right\}, \label{eq:impute_model}
\end{equation}
where $\hat{Y}_n(t) = \hat{f}_t(Y_n, \bm{z}_n, J_n; \hat{\bm{\theta}}_t)$ and $\MM:=\{f_0, f_1\}$ is used to emphasize the dependency of the estimator on the regression functions used. We summarize this procedure in \Cref{alg:GOBE}.
\begin{algorithm}
\caption{Generalized Oaxaca-Blinder Estimators}\label{alg:GOBE}
\begin{algorithmic}
\Require Data $\{ (Y_n, J_n, \covariate_n) \}_{n=1}^N$, regression model $\M = \{f_0, f_1\}$, where $f_0$ is the regression model for $T_0$ and $f_1$  for $T_1$.
\State For $t\in\{0,1\}$, and $\DD_t:=\{(Y_n, \bm{z}_n)\}_{n\in\II_t}$, fit regression model $f_t$:
\[    
	\fittedparam{t} \in \arg\min_{\bm{\theta}}\mathcal{L}\left\{\DD_t;\bm{\theta}\right\}.
\]
\State For $n=1,\ldots,N$, impute values 
    \[
        \hat{Y}_n(t) = \hat{f}_t(Y_n, \covariate_n, J_n;\fittedparam{t}) = \begin{cases} Y_{n} &\mbox{ if } t = J_n \\
        \hat{f}_{t}(Y_n, \bm{z}_n, J_n; \hat{\bm{\theta}}_t)&\mbox{ if } t = 1-J_n.\end{cases}
    \]
\State For $t = 0, 1$ estimate the the mean-squared error of the model:
\[
    \hMSE_{\MM, t} = \frac{1}{|\II_t|-1} \sum_{n \in \II_t} \left\{ Y_n - \hat{Y}_n(t) \right\}^2,
\]
and
\[
	 \hVAR_{\M}:=\frac{\hMSE_{\MM, 1}}{|\II_1|} + \frac{\hMSE_{\MM, 0}}{|\II_0|}.
\]
\State \Return Estimate and corresponding confidence intervals $\hatATE_{\M}$ as per \Cref{eq:impute_model}
    and
	\[
            \hCI_{\M}(\alpha):=\hatATE_{\M} \pm z_{1-\frac{\alpha}{2}}\sqrt{\hVAR_{\M}},
	\]
    where $z_{\alpha}$ is the $100\times \alpha \%$ quantile of the cumulative density function [CDF] of the standard normal distribution.
\end{algorithmic} 
\end{algorithm}

The Gaussian assumption on the confidence intervals returned by \Cref{alg:GOBE} is asymptotically justified under mild conditions for large classes of models (see \citet[Theorem 4]{guo2021generalized}).

\paragraph{Difference-in-means as GOBE} Notice that the difference-in-mean estimator introduced in \Cref{eq:unATE} can be viewed as a generalized Oaxaca-Blinder estimator. Indeed $\unATE$ satisfies \Cref{eq:impute_model} for the choice $\hat{f}_t(Y_n,\covariate_n, J_n;\fittedparam{t}) = \frac{1}{|\II_t|}\sum_{n' \in \II_t} Y_{n'}$, where we impute the missing values with the group mean, irrespective of the value of the covariates.

\paragraph{Linear regression as GOBE} The standard linear-regression adjusted estimator is a GOBE. This estimator is obtained by fitting via ordinary least squares [OLS] the following regression:
\begin{equation}
    Y_n \sim \beta_0 + \beta_1 J_n + \bm{\gamma}^\top \covariate_n + \bm{\delta}^\top J_n(\covariate_n - \bar{\covariate}), 
    \label{eq:intATE_reg}
\end{equation}
and using the estimate $\intATE:=\hat{\beta}_1$.
Here $\bar{\covariate} \in \RE^K$ is the average for each of the $K$ components, computed across the $N$ units. \citet{lin2013agnostic} shows that asymptotically $\intATE$ is unbiased and has smaller variance than $\unATE$. To see $\intATE$ as a ``GOBE'', notice that  it can be equivalently obtained by fitting via OLS in {two separate linear regressions}: for $n \in \II_t$, $t \in \{0,1\}$, fit $Y_n \sim \bm{\theta}^\top_t(\covariate_n - \bar{\covariate})$. We recover the GOBE formulation of \Cref{eq:impute_model} by letting
\begin{equation}
    \hat{f}_t(Y_n, \covariate_n, J_n; \fittedparam{t}) = \begin{cases} Y_n &\mbox{ if } t = J_n \\ 
    \fittedparam{t}^\top \covariate_n &\mbox{ if } t \neq J_n.
    \end{cases} \label{eq:predict_LR}
\end{equation}
See, e.g. \citet[Lemma 3]{lin2013agnostic} for a proof of why \Cref{eq:predict_LR} and \Cref{eq:intATE_reg} lead to the same estimator.


One can also employ general non-linear regression models to perform adjustments. \citet{guo2021generalized} provide  conditions  under which regression models produce unbiased and asymptotically normal estimates, justifying the Gaussian approximation in \Cref{alg:GOBE}. 
Building on these results, \citet{cohen2020no} propose a two-step GOBE. First a GOBE is fitted, and then a second GOBE with a linear regression model using imputed values $x_n := \hat{f}_{1-J_n}(Y_n,\covariate_n,J_n;\fittedparam{1-J_n})$ as the only covariate for each outcome $Y_n$ is used to produce the final estimate. This ``two-step'' GOBE is asymptotically unbiased, normally distributed, and more efficient than the difference-in-means estimator. See  \citet{guo2021machine,jin2021towards} for other recent approaches to develop flexible models in online A/B testing.


\section{Implementing GOBEs at scale} \label{sec:complexity}

As already discussed in \Cref{sec:intro}, drawing conclusions from online A/B tests can be challenging: experiments often consist of small changes related to details of the user experience. Consequently, associated effect sizes can be very small and hard to detect even in large samples. Despite being small, these can lead to large downstream impacts. Exactly because of this reason, employing models that allow for precise estimates of the causal effects is important: more precise estimates of the effects of the interventions can allow practitioners to detect smaller effect sizes and crucially shorten the experimentation time needed in order to obtain a conclusive answer about the effectiveness of a treatment.

Ultimately, it would be desirable to have an end-to-end automated inference engine which produces, for each experiment, the ``best'' possible estimate for the causal effect under study, without requiring experimenters to specify which model and covariates should be employed for this task. In practice, assessing which estimator is best is far from being trivial. Indeed, while on the one hand the idea of developing ad-hoc large models with curated covariates for an individual experiment of interest seems appealing for variance reduction, on the other hand large-scale causal inference engines have to rely on estimators that perform well on average across all experiments. That is, the methods used need to be:
\begin{itemize}
    \item Scalable: companies typically run a very large number of experiments every year, and their computational resources are limited. It is undesirable for practitioners to have to wait for their results due to long analysis run times (e.g., to solve the minimization problem in \Cref{alg:GOBE}).
    \item Reliable: the team maintaining the infrastructure is often small relative to the customer base it serves. The methods implemented need to rely on algorithmically sound routines that produce stable estimates of the causal effects of interest.
    \item Interpretable: the results of the experiments are used by practitioners for policy-making. It is therefore imperative that the estimates produced are transparent, easy to interpret, and do not require specialized knowledge.
\end{itemize}

Because of these reasons, in our experiments presented in \Cref{sec:experiments} we only employ linear models, their regularized counterparts (LASSO, ElasticNet, Ridge and principal components regression), and one simple instance of a generalized linear model. 
Extending our analysis to more complicated models, and assessing their feasibility in a production setting is part of ongoing investigations.

We here describe in detail the regression models we fit to experimental data to benchmark the performances of different GOBE estimators. 
We have already discussed the difference in means [DIM] and simple linear regression [LR] estimators, and their characterization as GOBEs in \Cref{sec:gobes}. 
Ridge regression, LASSO and elastic net are extremely popular ``regularized'' counterparts of simple linear regression model, in which the weight vector $\bm{\theta}$ is ``shrunk'' using a penalty. Formally, given a regularization parameter $\gamma>0$, we minimize with respect to $\bm{\theta}$ the loss function
\[
	\mathcal{L}(\DD_t;\bm{\theta}):= \sum_{n \in \II_t} \left( y_n - \bm{\theta}^\top \covariate_n\right)^2 + {\gamma} \| \bm{\theta}\|_\ell^2,
\]
where $\ell=1$ for LASSO and $\ell=2$ for Ridge regression  (i.e., regularize using the $\ell$-1 or $\ell$-2 norm). Elastic net regression is obtained by combining the $\ell_1$ and $\ell_2$ penalties on the regression coefficients. Formally, in this case, we minimize the loss function
\[
	\mathcal{L}(\DD_t;\bm{\theta}):= \sum_{n \in \II_t} \left( y_n - \bm{\theta}^\top \covariate_n\right)^2 + \gamma  \lambda \sum_{k=1}^K |\theta_{t,k}|
+ \frac{\gamma  (1 - \lambda)}{2}  \sum_{k=1}^K \theta_{t,k}^2.
\]
Here $\lambda$ trades off the importance of the $\ell_1$ and $\ell_2$ penalties. Differently from linear regression and the simple difference in means, these regularized models crucially depend on the tuning of some regularization hyperparameter. Towards the goal of having a streamlined, automated procedure to fit these models, we adopt a standard cross-validation approach.
For each model, we repeatedly minimize the objective function across a predetermined number of different values of the regularization parameters. For each of these values, we split the data in $5$ random folds, and fit the model $5$ times by iteratively leaving out one fold of the data. For each fold, we compute the coefficient of determination on the left-out-data using the fitted coefficient and choose the optimal regularization level by picking the value that achieved the maximum average coefficient of determination across the folds, and re-fit the model using the full dataset. 
We also consider principal component regression [PCR] --- where we first reduce the dimensionality of the regressors using their projections onto principal components, and then use these as covariates in a linear regression, as well as an instance of a generalized linear model using a Tweedie distribution kernel.

\section{Experiments} \label{sec:experiments} \label{sec:exp}

\subsection{Data description}

\begin{figure}
    \centering
    \includegraphics{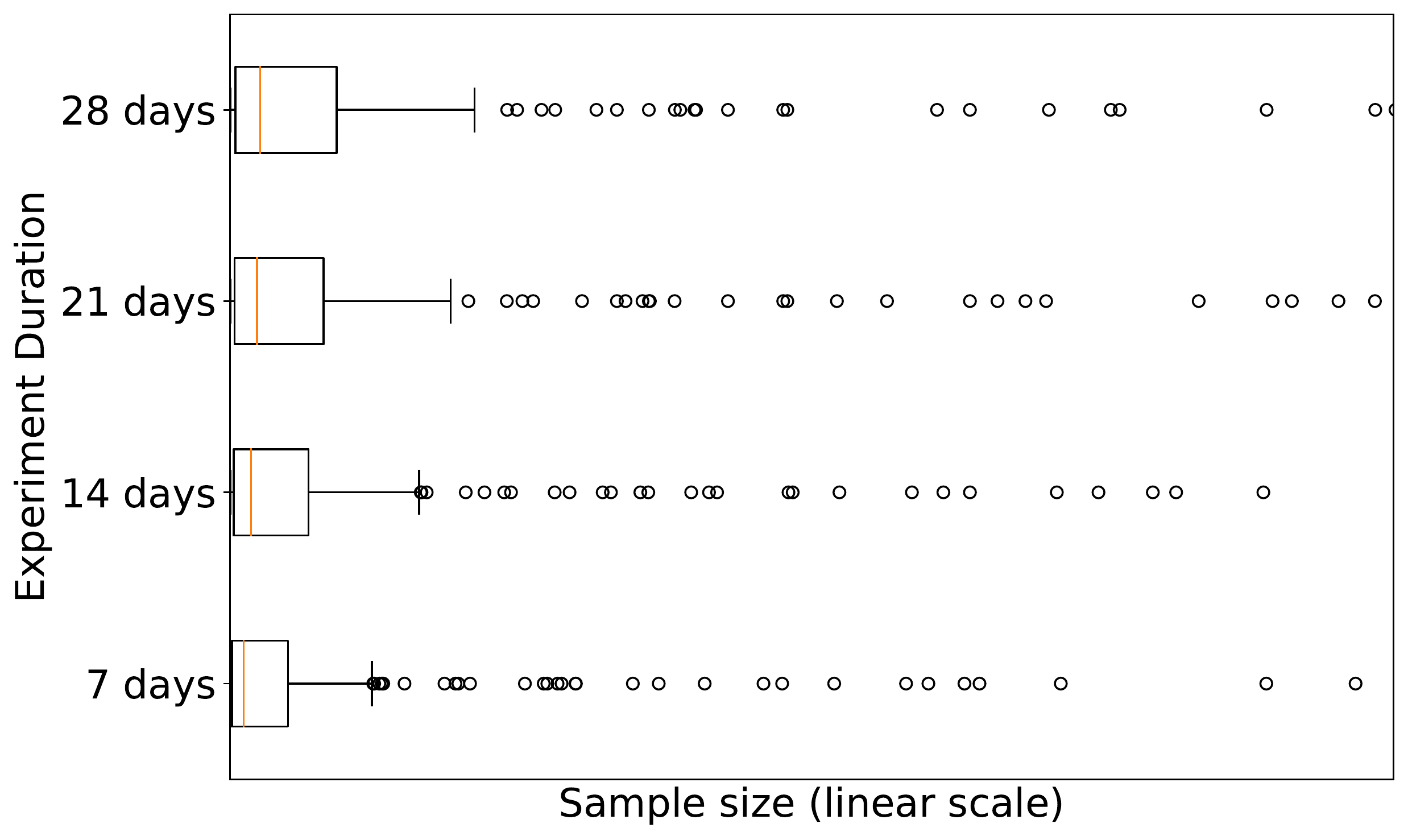}
    \caption{Empirical distribution of the sample sizes of the experiments considered in the meta-analysis.}
    \label{fig:sample_size_distribution}
\end{figure}

For our experiments, we consider a representative set of $W=100$ A/B tests.
These have been running in production over the course of the last two years, at different times of the year. Each experiment corresponds to a different intervention.
For simplicity, in our analysis we only consider one pairwise comparison per experiment ($T_1$ versus $T_0$) --- even though some experiments might have more than two treatment arms.
For each experiment, we run our data analysis pipeline and compute estimates of the causal effects after collecting data for a total time of $D \in \{7, 14, 21, 28\}$ days. 
For any analysis duration, the sizes of the experiments (total number of customers in the $T_1$ and $T_0$ arms) varies considerably (\Cref{fig:sample_size_distribution}). 

For all these experiments, we track the same KPI of interest. Since the scale of this KPI varies across experiments, in our illustrations and analysis we focus on the percent ATE (or lift), which is defined as $\LLIFT:=\ATE/|\bar{Y}_0|$.
 We use as covariates less than 10 pre-exposure values of customer metrics correlated with the KPI.
 
 \subsection{Variance reduction} \label{sec:vr}

Intuitively, a more precise estimator (with estimated lower variance) leads to better estimates, and directly translate in faster and better decision making (as we further discuss in \Cref{sec:duration_recommendation}). 
We estimate the performance of model $\MM$ in terms of precision by computing their estimated percentage variance reduction $\hVR_{\MM}$ with respect to the baseline $\DIM$. Recalling that $\hVAR_{\M}$ is the estimated variance of the $\ATE$ under model $\MM$ (as per \Cref{alg:GOBE}), we define:
\[
	\hVR_{\M}:= 100 \times \left\{ 1 - \frac{\hVAR_{\M}}{\hVAR_{\mathrm{DIM}}}\right\}.
\]

\begin{figure}
    \centering
    \includegraphics{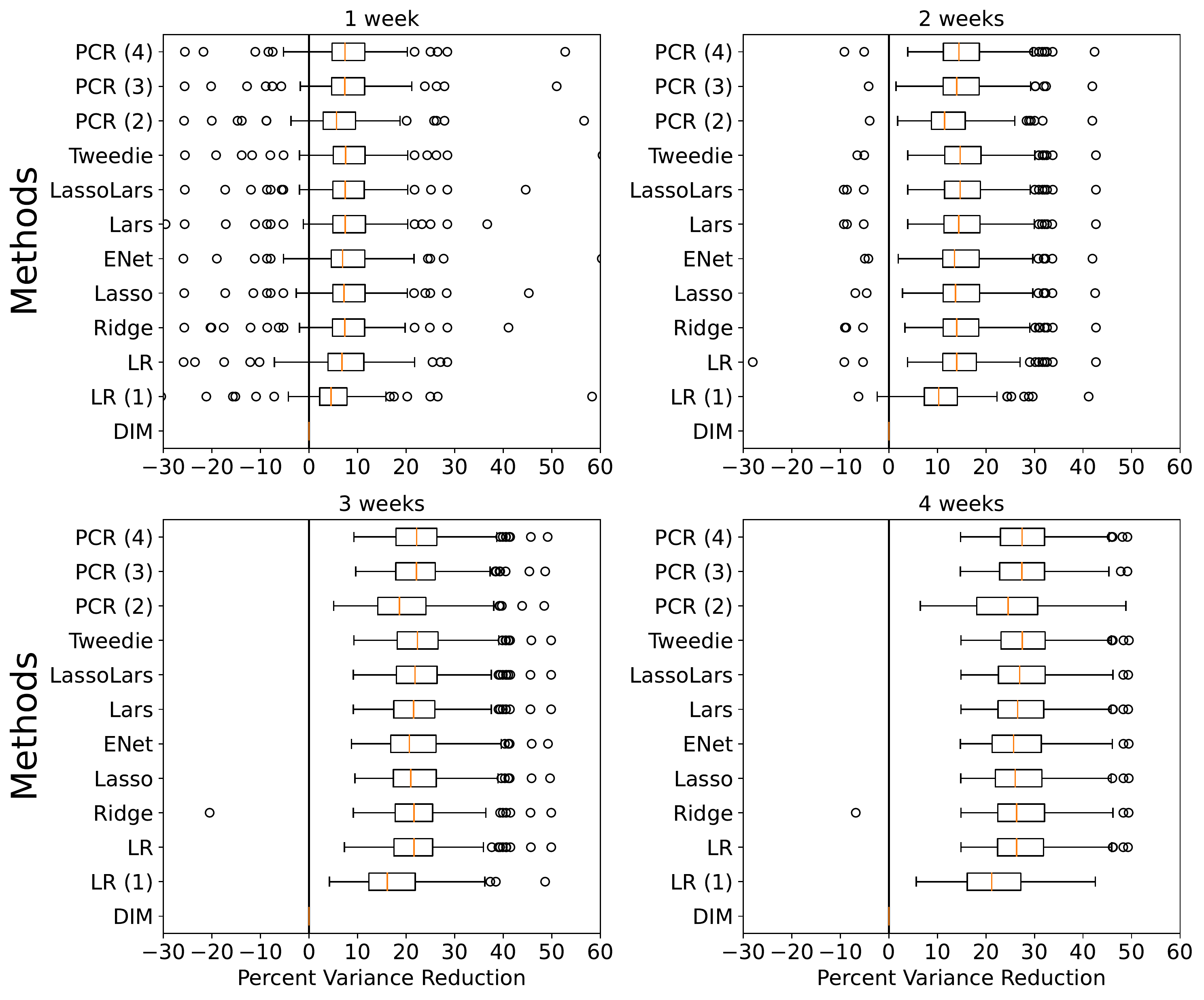}
    \caption{Boxplots of the estimated variance reduction across the experiments considered in the analysis. 
    Each subplot in the figure corresponds to a different analysis duration time, and each row in the boxplot refers to a different model $\M$.}
    \label{fig:variance_reduction_all_weeks}
\end{figure}

We plot in \Cref{fig:variance_reduction_all_weeks} the variance reduction as a function of the duration of the analysis across all the experiments in the meta-analysis. Three main findings emerge:
\begin{itemize} 
	\item As expected (see, e.g. \citet[Theorem 4]{guo2021generalized}), covariate adjusted estimates generally have smaller variances. 
	\item Larger variance reduction is observed in longer analyses, which are characterized by more stable customer behavior. 
	\item The performance observed across different covariate adjusted methods is similar. 
	We analyze in \Cref{sec:computation} the performance and computation cost of these methods in relation with the number of (potentially noisy) regressors.
\end{itemize}

Next, we further try to understand the relationship  between analysis duration, sample size and model precision.
For a given duration of the analyses (e.g., $D= 7$ days), let $F_{N,D}:\NN\to[0,1]$ be the empirical cumulative density function [CDF] of the sample size $N$ of the experiments after $D$ days, and let $F^{-1}_{N,D}:[0,1]\to \NN$ be its inverse. E.g., $F^{-1}_{N,7}(0.6)$ is the sample size of the 60\%-largest experiment amongst the 7-day analyses.  
We consider the variance reduction gains within the first quartile (experiments with sample size $N \in [F_{N,D}^{-1}(0),F_{N,D}^{-1}(0.25)$) and in the last quartile (experiments with sample size $N \in [F_{N,D}^{-1}(0.75),F_{N,D}^{-1}(1)$). We observe different behaviors at duration $D=7$ and $D=28$. Specifically, for the shorter analysis time ($D=7$ days) variance reduction is particularly evident in \emph{larger} experiments. However, for $D=28$ smaller experiments seem to be benefitting the most from covariate adjustments. More broadly, we expect the variance reduction induced by covariate adjustments can vary with the experiment size and duration, and the choice of covariates and model used. We advise practitioners to extensively analyze their data, before the experiment is run, prior to adopting a regression model.

\begin{figure}
    \centering
    \includegraphics{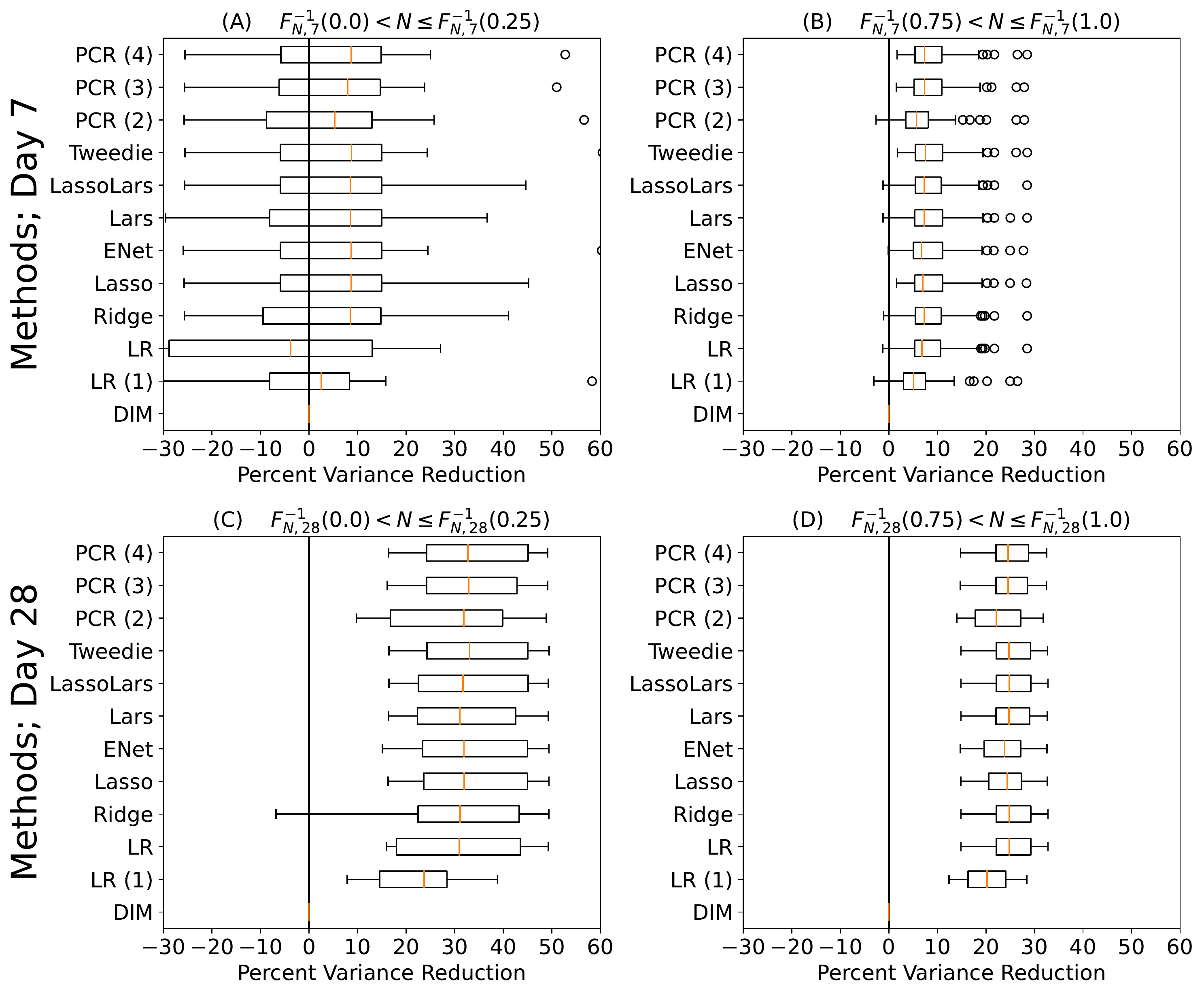}
    \caption{Boxplots of the estimated variance reduction. 
    First row: 7-day analyses. Second row: 28-day analyses. Left subplots (A, C): $\hVR_{\MM}$ in the smallest 25 experiments. 
    Right subplots (B, D) $\hVR_{\MM}$ for the largest 25 experiments. 
    Each row in each boxplot corresponds to a different model $\M$.}
	\label{fig:variance_reduction_by_duration_and_size}
\end{figure}

\subsection{Robustness to chance imbalance} \label{sec:chance_imbalance}

In an A/B test, treatment arms should be \emph{ex-ante} comparable. 
That is, by virtue of the randomization, the distribution of the covariates for the units in the treatment and control group should coincide. 
This is not only a property of a correctly constructed A/B test, but also a fundamental requirement that such an experiment should satisfy to yield valid inferences. 
Consider an experiment in which the value of a given covariate $x$ is predictive of the outcome $y$ (e.g., units with higher $x$ tend to have higher $y$). 
If the triggering logic systematically allocates with higher (or lower) probability units with higher value of $x$ to the treatment, condition (C1) in \Cref{sec:gobes} is violated. 
As a consequence, inferences obtained from the A/B test are going to be invalid.

\begin{figure}
    \centering
    \includegraphics{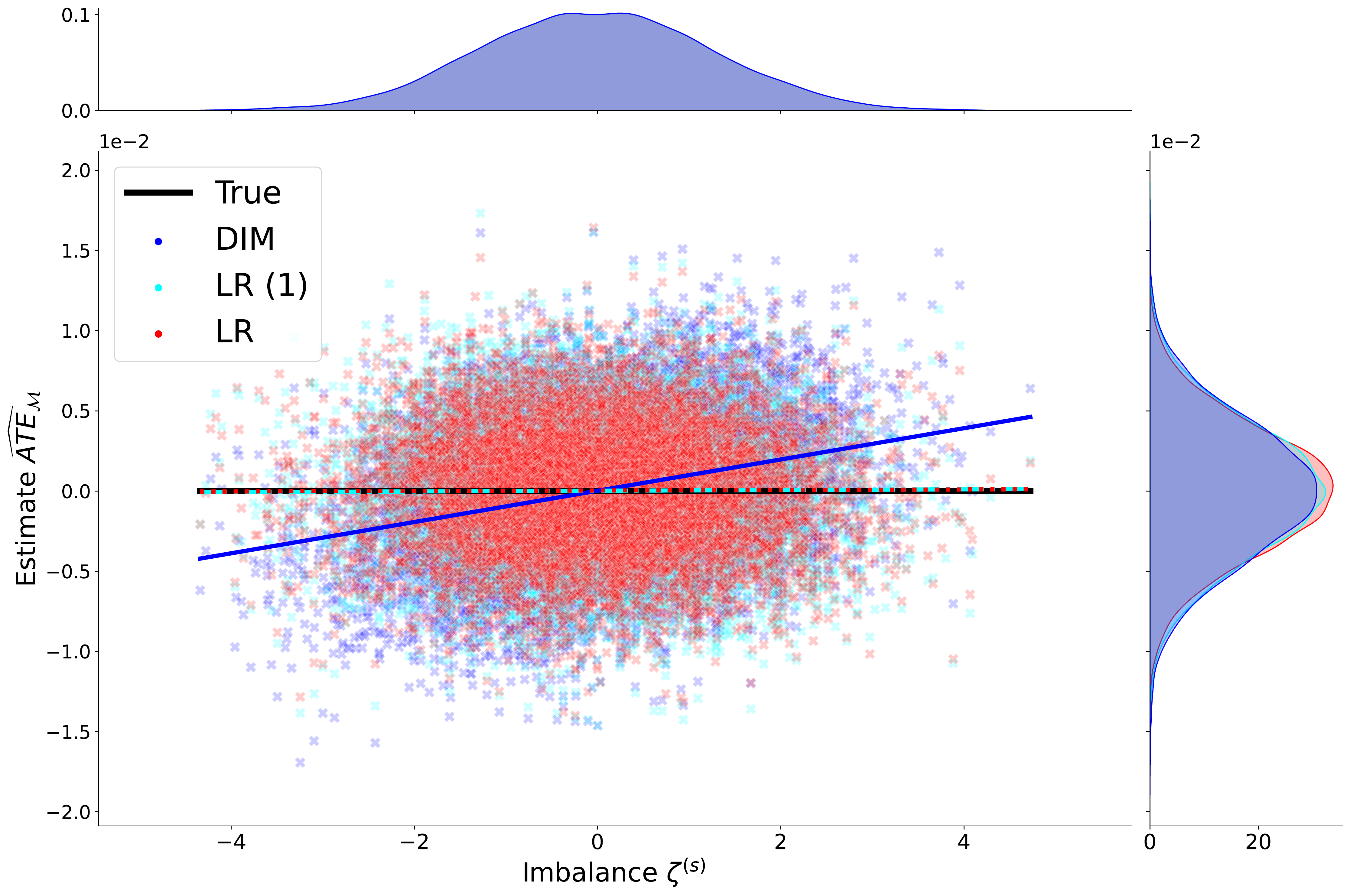}
    \caption{Comparing DIM and LR estimates for a single experiment for $S = 10,000$ re-randomizations. Each dot corresponds to a different $(\zeta^{(s)}, \hatATE_{\MM})$ combination.}
    \label{fig:imbalance}
\end{figure}

Even for experiments in which the triggering logic determining treatment assignments is correctly specified, however, it can be the case that in practice an experiment leads to imbalanced treatment arms. 
E.g., an experiment in which the triggering logic follows \Cref{eq:triggering} can result in an ``unlucky'' split of the data, in which the covariate values in treatment arms are not comparable. 
Concretely, assignment variables $J_{1:N}:=\{J_1,\ldots,J_N\}$ could define a control group $\II_0$ containing units that have on average much higher values of the KPI of interest in the pre-experimental period with respect to $\II_1$ (or \emph{vice versa}). In the presence of high pre-experimental covariate imbalance, practitioners worry whether they can trust their findings. We here empirically show the following:
\begin{itemize}
	\item Under high imbalance, the difference in means estimator can systematically lead to wrong conclusions. 
	In other terms: $\hatATE_{\DIM}$ is unbiased unconditionally on the covariate imbalance, but it can be conditionally biased. See \Cref{fig:imbalance}.
	\item Covariate adjusted methods alleviate this concern, and are robust to pre-experimental chance imbalance. See \Cref{fig:robustness}.
\end{itemize}

Notice: in this section, we focus on simple linear regression --- qualitative findings for other methods are similar and omitted. To get us started, we need an operational definition of imbalance to quantify the pre-experimental comparability of the control and treatment arms. Let $x_n$ denote the pre-experimental value of the KPI of interest $Y_n$ for unit $n$, and define the imbalance parameter $\zeta$:
\begin{equation}
	\zeta:=\zeta(x_{1:N}, J_{1:N}) = \bar{x}(1) - \bar{x}(0), \label{eq:imbalance}
\end{equation}
where $\bar{x}(t) :=  \sum_{n=1}^N x_n 1(J_n = t) /  \{ \sum_{n=1}^N 1(J_n = t) \}$ is the average value of the KPI in the pre-experimental period for units later exposed to treatment arm $t$.
Intuitively, when $|\zeta|$ is large, the two groups (control and treatment) are not ex-ante comparable. 
When the imbalance is sufficiently severe, experimenters worry that the estimates might not be trustworthy. 
In turn, this typically leads to the necessity of re-randomizing the experiment. 
This causes inefficiency in the experimentation pipeline: re-randomizing is expensive as it requires using additional computational resources, and postponing launch decisions. 
Methods whose inferences are less sensitive to randomization bias are therefore preferable.

We now show on our real data that using a covariate adjusted estimator can lead to substantially better results than the unadjusted estimator, even in the presence of severe imbalance. In turn, this reduces the need to re-randomize and experimentation cost.

\subsubsection{A/A analysis} \label{sec:validation_AA}

To study robustness of covariate adjustments to chance imbalance, we here adopt the following ``A/A'' approach. Given an experiment of interest, we restrict our attention to a single arm in the experiment (e.g., control). Namely, we only consider the subset of the units $n\in\II_t$ exposed to policy $t$, for a fixed $t$, and discard all the other units, together with their covariates. For notation simplicity, in what follows we denote $\II:=\II_t$ the set of units in treatment arm $t$. After this pre-processing, we can treat the  units in $\II$ as if they were obtained from an A/A test. That is, if we were to randomly split them into two ``fake'' treatment arms, we would have by construction that the ground truth average causal effect is known and equal to $0$. A similar experimental setup is adopted e.g.\ in \citet[Section 4]{guo2021generalized}.

\subsubsection{Monte Carlo simulation} To assess robustness to imbalance, we adopt a Monte Carlo approach. We fix a large integer $S$ and for each $s=1,\ldots,S$, we  randomly split $\II$ into two groups, creating A/A re-randomization groups $\II_{0}^{(s)}, \II_{1}^{(s)}$ such that $\II_{0}^{(s)} \cup \II_{1}^{(s)} = \II$ and $\II_{0}^{(s)} \cap \II_{1}^{(s)} = \varnothing$. We then define the A/A arm indicator $J_{n}^{(s)} := 1(n \in  \II_{1}^{(s)})$, and use \Cref{alg:GOBE} to fit $\hatATE^{(s)}_{\MM}$ using data $\mathcal{D}_t^{(s)} = \{Y_{\II}, J_{\II}^{(s)}, X_{\II}\}$ for all the estimators $\MM$ under consideration. Here $Y_{\II} = \{Y_n\;:\; J_n \in \II\}$. That is, we fit $\hatATE$ assuming that the units in control are those with index in $\II_{0}^{(s)}$, and the unit in treatment are indexed by $\II_{1}^{(s)}$. Importantly, notice that for every re-randomization $s$ we induce a split-specific level of imbalance $\zeta^{(s)}$ as per \Cref{eq:imbalance}. Moreover, by construction the ATE is $0$, since we here let the outcome $Y_n$ be fixed, regardless of the value of $J_{n}^{(s)}$. We summarize this procedure in \Cref{alg:AA}.
\begin{algorithm}
\caption{A/A test}\label{alg:AA}
\begin{algorithmic}
\Require Data $\mathcal{D}:=\{(y_n, \bm{z}_n)\}_{n\in\II}$, set of regression models $\mathfrak{M} = \{\MM_1, \ldots, \MM_W\}$, treatment arm $t$.
\State Let $\II := \{n \in [N] \;:\; J_n = t\}$.
\For{$s = 1, \ldots, S$}
	\State Split $\II$ into $\II_{0}^{(s)}, \II_{1}^{(s)}$ at random such that $\II_{0}^{(s)} \cup \II_{1}^{(s)} = \II$ 
	and $\II_{0}^{(s)} \cap \II_{1}^{(s)} = \varnothing$. Let $J_{n}^{(s)} := 1(n \in  \II_{1}^{(s)})$.
	\State Let $\bar{x}(\II_t^{(s)}) ~:=~ \sum_{n\in \II_t^{(s)}} x_n /|\II_t^{(s)}|$ and compute 
		\[
			\zeta^{(s)} = \bar{x}(\II_{1}^{(s)}) - \bar{x}(\II_{0}^{(s)}).
		\]
	\For{$\MM \in \mathfrak{M}$}
		\State With $\mathcal{D}^{(s)} ~:=~ \{Y_{\II}, J_{\II}^{(s)}, X_{\II}\}$, estimate 
			$\hatATE_{\MM}^{(s)}, \hCI_{\MM}^{(s)}(\alpha)$ using \Cref{alg:GOBE}.
	\EndFor
\EndFor
\end{algorithmic}
\end{algorithm}

\subsubsection{Validation} Once we have performed the Monte Carlo procedure described above, we have access to an empirical bivariate distribution of the estimator $\hatATE_{\MM}$ as a function of the imbalance level $\zeta$. We provide a visualization of how different estimators perform in \Cref{fig:imbalance}, where  \Cref{alg:AA} has been run on the control arm of a single experiment in the meta-analysis for $S=10{,}000$. We make a scatterplot of the imbalance level $\zeta$ (horizontal axis) against $\hatATE_{\MM}$ for $\MM \in \{\DIM, \LR 1, \LR\}$ (vertical axis). Here $\LR 1$ signifies that we regress the KPI only against its pre-experimental value, while $\LR$ uses all the available covariates. We plot on the top the marginal distribution of the imbalance level $\zeta$ (which is as expected centered around $0$), and on the right the marginal density of each estimator $\hatATE_{\MM}$. We also plot a solid color line that is the linear fit of $\hatATE_{\MM}$ agains $\zeta$. It is evident from \Cref{fig:imbalance} that $\hatATE_{\DIM}$ is unconditionally unbiased (its expectation over re-randomization $s=1,\ldots,S$ coincides with the true ATE $0$), but it is not unbiased conditionally on imbalance. 

Because the true value of the true underlying average causal effect is known, this information directly translates in a joint distribution for the accuracy of the estimator as a function of the imbalance. Estimators that are less sensitive to the imbalance allow experimenters to be confident about the results obtained even when such imbalance is present. 
\emph{Vice versa}, an estimator that is sensitive to the imbalance level will make experimenters doubt their findings when the pre-experimental covariates are imbalanced. 
In turn, stable estimators will result in more efficient experimentation pipelines, in which data from ``unlucky splits'' are still useful to draw conclusions about the causal effect of interest. 

We now introduce a number of metrics that allow us to translate this intuition into a quantitative assessment of the quality of the estimator as a function of the imbalance level.
 Fix a value $\kappa \in \NN$, and  create index sets $\ES_1,\ldots,\ES_\kappa$, where for each $j=1,\ldots, \kappa$, the index set $\ES_j \subset \{1,\ldots,S\}$ 
	contains the indices associated with the values of $\zeta^{(s)}$ within the $100\times\frac{j-1}{\kappa}\%$ to the $100\times\frac{j}{\kappa}\%$ quantile of the empirical distribution of $\zeta^{(s)}$. 
	Then, we partition the values $(\hatATE_{\MM}^{(s)}, \zeta^{(s)})$ into $\kappa$ splits of equal size ($\kappa$-iles), according to the (sorted) value of $\zeta^{(s)}$.
	For each $j = 1,\ldots,\kappa$, let $G_{\MM, j}:\RE\to[0,1]$ be the empirical cumulative density function of the estimates $\hatATE_{\MM}^{(s)}$ falling into the $j$-th bucket, and let $G^{-1}_{\MM, j}:[0,1]\to\RE$ be its inverse (e.g., for $\kappa=10$, we obtain the median value of the estimated ATEs which fall between the 30th and 40th percentile via $G^{-1}_{3,\MM}(0.5)$).
	 Within each of the $\kappa$ buckets, we compute
	\begin{itemize}
		\item the estimated MSE in the $j$-th bucket:
			\begin{equation}
				\hMSE_{\MM,j} = |\ES_j|^{-1} \sum_{s\in \ES_j} \left(\hatATE_{\MM}^{(s)} - \ATE \right)^2. \label{eq:hMSE}
			\end{equation}
		\item the square distance of the median value of $\hatATE_{\MM}^{(s)}$ in the $j$-th bucket to the true value $\ATE$:
			\begin{equation}
				\hmediandist_{\MM,j} = (G^{-1}_{\MM, j}(1/2)-\ATE)^2. \label{eq:hmediandist}
			\end{equation}
		\item the excess fraction of $\hatATE_{\MM}^{(s)}$ in the $j$-th bucket which underestimate or overestimate the true effect $\ATE$:	
			\begin{equation}
				\hexcessfrac_{\MM,j} = \frac{\max\{ \hat{q}_{\MM,j}^+, 1-\hat{q}_{\MM,j}^-\} - 1/2}{1/2}, \label{eq:hexcessfrac}
			\end{equation}
		where $\hat{q}_{\MM,j}^+ =  \arg\min_\alpha \{G^{-1}_{\MM,j}(\alpha)\ge \ATE\}$ is the quantile associated with the smallest value $\hatATE_{\MM}^{(s)}$ 
		in the $j$-th bucket to be above $\ATE$   
		and $\hat{q}_{\MM,j}^-=  \arg\max_\alpha \{G_{\MM,j}^{-1}(\alpha)\le \ATE\})$ is the quantile associated with the largest value $\hatATE_{\MM}^{(s)}$ in the $j$-th bucket to be 
		below the true $\ATE$. Notice: when the distribution of the estimates is centered around the true value, \Cref{eq:hexcessfrac} is close to $0$. 
		In the presence of large conditional bias, it approaches $1$.
	\end{itemize}
\begin{figure}
    \centering
    \includegraphics{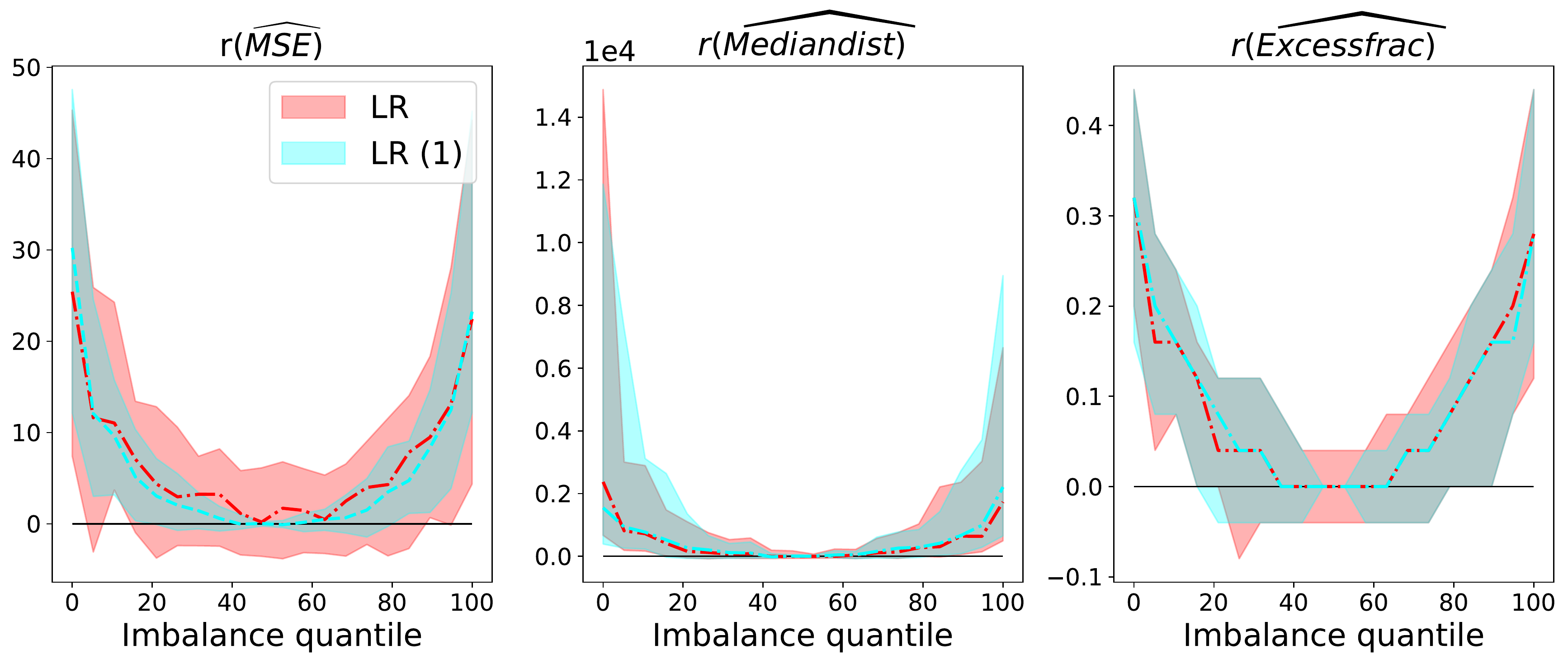}
    \caption{Summaries of the relative counterparts (\Cref{eq:hrel}) of the robustness metrics introduced in \Cref{eq:hMSE,eq:hmediandist,eq:hexcessfrac} across the $M$ experiments at day $7$ of the analysis. Higher values correspond to higher sensitivity of $\DIM$ relative to $\LR, \LR 1$. Solid line: median across experiments; Shaded regions : (25\% -- 75\%) percentiles. }
    \label{fig:robustness}
\end{figure}

\begin{figure}
    \centering
    \includegraphics{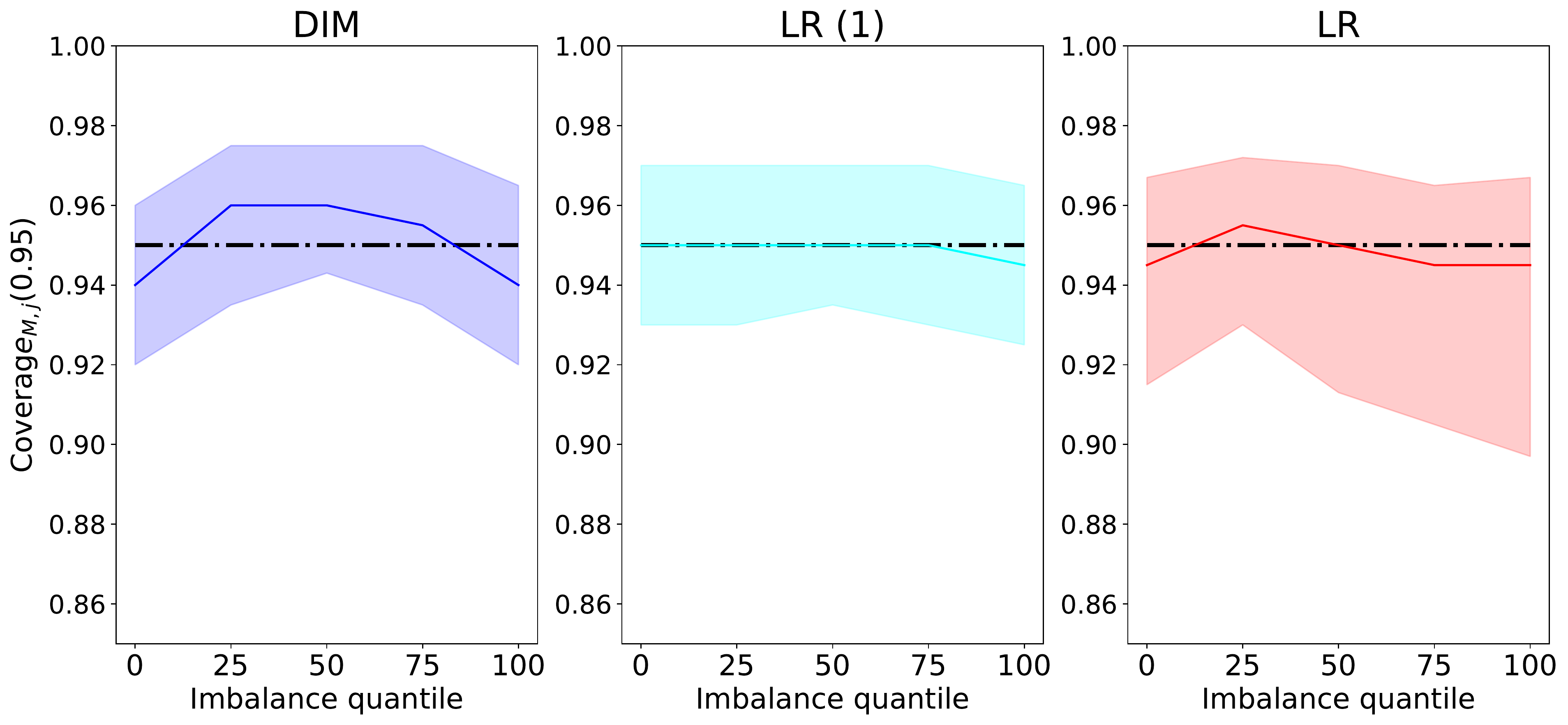}
    \caption{Summaries of $\hcoverage_{\MM,j}(\alpha)$ for $\MM \in \{\DIM, \LR 1, \LR\}$ (vertical axis, left to right), at $\alpha = 95\%$ across $\kappa = 5$ buckets (horizontal axis). The solid line tracks the median coverage across the $M$ experiments, and the shaded regions cover 10\% -- 90\% percentiles across these experiments. Results are relative to the analysis at day $D=7$. The solid black line is the target nominal value 95\%.}
    \label{fig:calibration}
\end{figure}

Since these three metrics can all be regarded as notions of loss (the lower the value, the better), and since we're interested in assessing whether a covariate adjusted method $\MM$ achieves lower loss than the default unadjusted method $\DIM$, we also define for a method $\MM$ their ``relative'' (to $\DIM$) counterpart as 
\begin{equation}
	r(\hmetric_{\MM,j}):=\frac{\hmetric_{\DIM,j}-\hmetric_{\MM,j}}{\hmetric_{\DIM,j}}, \label{eq:hrel}
\end{equation}
for $\hmetric \in \{\hMSE, \hmediandist, \hexcessfrac\}$. 
For these relative metrics, larger values indicate higher sensitivity of the difference in means estimator $\hatATE_{\DIM}$ with respect to an alternative covariate adjusted estimator  $\hatATE_{\MM}$  to the imbalance level $\zeta^{(s)}$. 
We report the value attained by these relative metrics across all the $W=100$ experiments on the control arm at analysis day $D=7$. 
Specifically, for each experiment we re-run \Cref{alg:AA} for $S=10{,}000$ re-randomizations and compute the relative metrics across $\kappa=20$ buckets (\Cref{fig:robustness}). We retain for each experiment and for each bucket the median value of the relative metric, and plot a solid line connecting the median (of these medians) across the $W$ experiments, for all imbalance quantiles $\ES_j$, for $j=1,\ldots,\kappa$, as well as a 50\% centered empirical intervals through the shaded region. We find a similar behavior across the metrics: they are close to zero when the imbalance $|\zeta^{(s)}|$ is small (i.e., around its median value across re-randomizations). As the absolute value of the imbalance level $|\zeta^{(s)}|$ increases, however, the value of the relative metrics also sharply increases, indicating larger sensitivity to the imbalance of the difference in means estimator with respect to the linear adjusted estimators (either using one or many covariates). We conclude by checking estimators' calibration (\Cref{fig:calibration}). 
Given width $\alpha = 0.95$, we compute the fraction of times that the estimated confidence interval spans the true value for each bucket $\ES_j$:
\begin{align}
	\hcoverage_{\MM,j}(\alpha)= \frac{\sum_{s\in \ES_j}\ind\left(\ATE \in \hatCI{\alpha}^{(s)}\right)}{|\ES_j|}.
\end{align}
Attaining the nominal coverage $\alpha$ means that the confidence intervals are well calibrated.
For this robustness metric the performance of the difference in means estimator is less less sensitive to pre-experimental imbalance than for the metrics considered in \Cref{fig:robustness}.

\subsection{Impact on experimentation time} \label{sec:duration_recommendation}

We next illustrate how smaller estimated variances can lead to shorter experimentation time. We consider a hypothesis testing framework, where $H_0$ is a null hypothesis of no effect of the treatment --- $H_0 : \{ \LIFT = 0 \}$ --- and $H_1$ is a fixed alternative  the treatment has a fixed percent effect of size $\delta$, $H_1 : \{\LIFT = \delta\}$. Based on the data collected so far (e.g., the first $D=7$ days of the analysis), we form a prediction on the number of future units that are going to trigger in the experiment as it progresses (adapting recent sample-size prediction methods, see  \citet{masoero2022more,richardson2022bayesian,camerlenghi2022scaled}). Given the hypotheses, the predictions, and the estimated variances at $D=7$, we compute the first future day $D'$ at which we expect to be able to reject the null hypothesis of no effect with at least $\pi = 80\%$ power under the alternative $H_1$ with a given significance $\alpha = 0.05$. As seen in \Cref{fig:sample_size_distribution}, higher precision (smaller variance) directly translates in shorter experimentation time. We see, again, very similar performance across the different covariate adjusted methods considered. 
\begin{figure}
    \centering
    \includegraphics{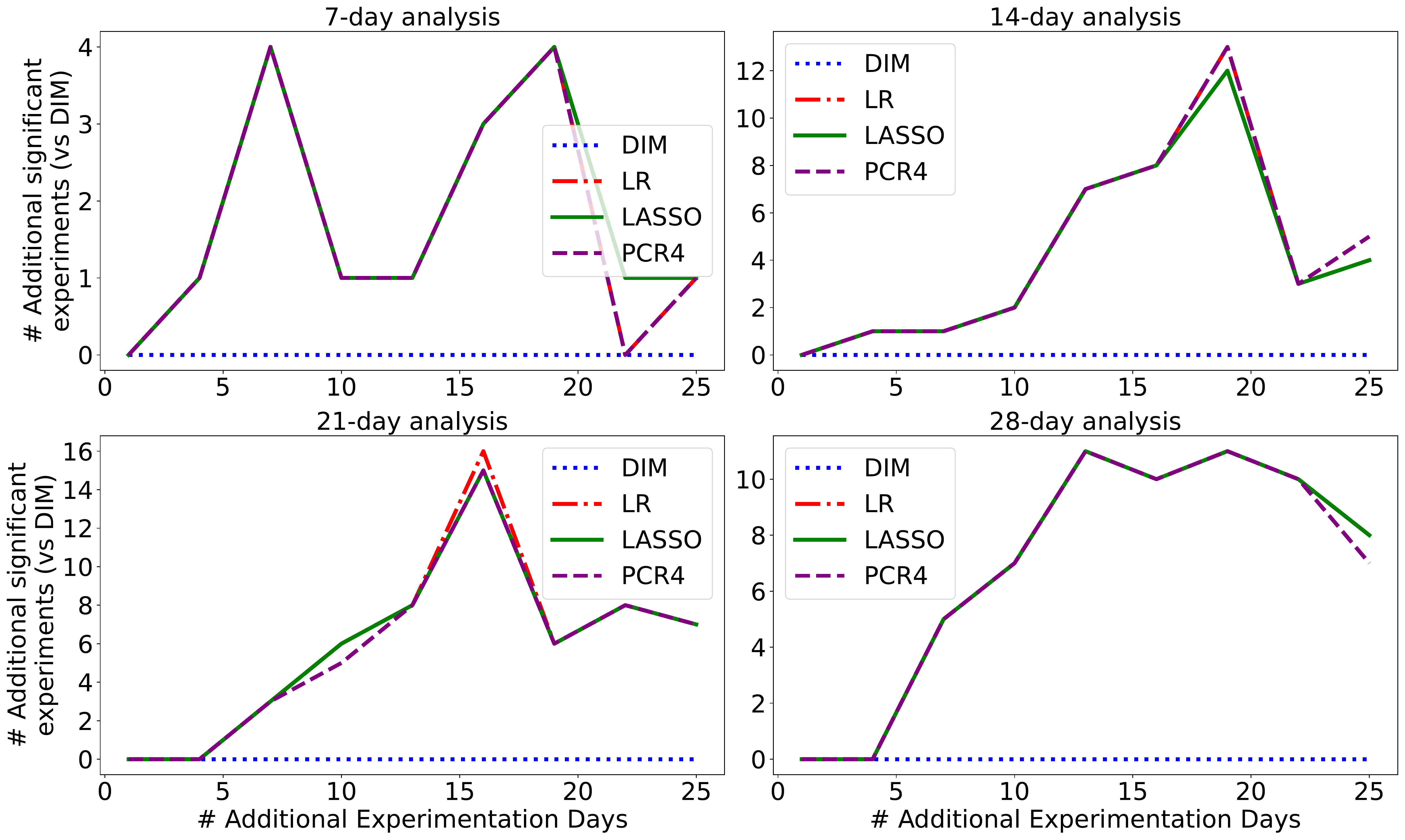}
    \caption{For a given number of additional experimentation days (horizontal axis), we plot the additional number of experiments (vertical axis) for which $H_0$ can be rejected at significance $\alpha=0.05$ with $80\%$ power given $H_1$ when the estimates are obtained using a model $\MM$ as opposed to the default $\DIM$. Different subplots refer to different experimental durations.}
    \label{fig:duration_rec}
\end{figure}

We emphasize that the predictions in \Cref{fig:duration_rec} depend on a number of factors: from the properties of the experiment (e.g., the observed means and variances of the KPI), to the choice of hypothesized fixed effect value $\delta$. However, the trend displayed in \Cref{fig:duration_rec} --- whereby smaller estimated variances translate in higher power and hence shorter duration of the experiments --- is expected: smaller estimated variances directly translate into shorter experiments.

\subsection{Making tradeoffs at scale: computation, robustness, interpretability} \label{sec:computation}

As already discussed in \Cref{sec:complexity}, large scale inference engines should be designed keeping in mind the constraints imposed by the scale at which they operate. We have already discussed how linear models and regularizations thereof are robust (e.g., to chance imbalance) and interpretable. We here analyze how the computation cost and the estimation accuracy scales with the number of (noisy) additional covariates. Specifically, we test how computation and accuracy are affected by augmenting the $K$ covariates with additional spurious covariates. To do so, we compute for each covariate $k$ the (empirical) first moment $\hat{\mu}_k$ and second moment $\hat{\sigma}^2_k$, and draw for each $n=1,\ldots,N$, a set of spurious covariates $\tilde{z}_{n,k} \sim \mathcal{N}(\hat{\mu}_k, \hat{\sigma}_k)$ i.i.d., for $k = 1,\ldots, K$. This produces an augmented set of covariates, $\tilde{\bm{z}}_n:=[z_{n,1},\ldots,z_{n,K},\tilde{z}_{n,1},\ldots,\tilde{z}_{n,K}]^\top$. In our experiments, we also consider even larger sets of covariates, obtained by drawing $L$ times from each kernel $\mathcal{N}(\hat{\mu}_k, \hat{\sigma}_k)$ for every $n=1,\ldots,N$. In the general case where we draw $L$ spurious folds, the covariates used are: 
\begin{align}
	\begin{split}
	\tilde{\bm{z}}_n:=[&\overbrace{z_{n,1},\ldots,z_{n,K}}^{\text{Real Covariates}},
				\overbrace{\tilde{z}_{n,1},\ldots,\tilde{z}_{n,K}}^{\text{First Spurious Fold}}, 
				\overbrace{\tilde{z}_{n,K+1},\ldots,\tilde{z}_{n,2K}}^{\text{Second Spurious Fold}}, \\
				&\ldots,\underbrace{\tilde{z}_{n,(L-1)K+1},\ldots,\tilde{z}_{n,LK}}_{L-\text{th Spurious Fold}}]^\top \in \RE^{(L+1)\times K}.
	\end{split}\label{eq:spurious}
\end{align}

\begin{figure}
    \centering
    \includegraphics{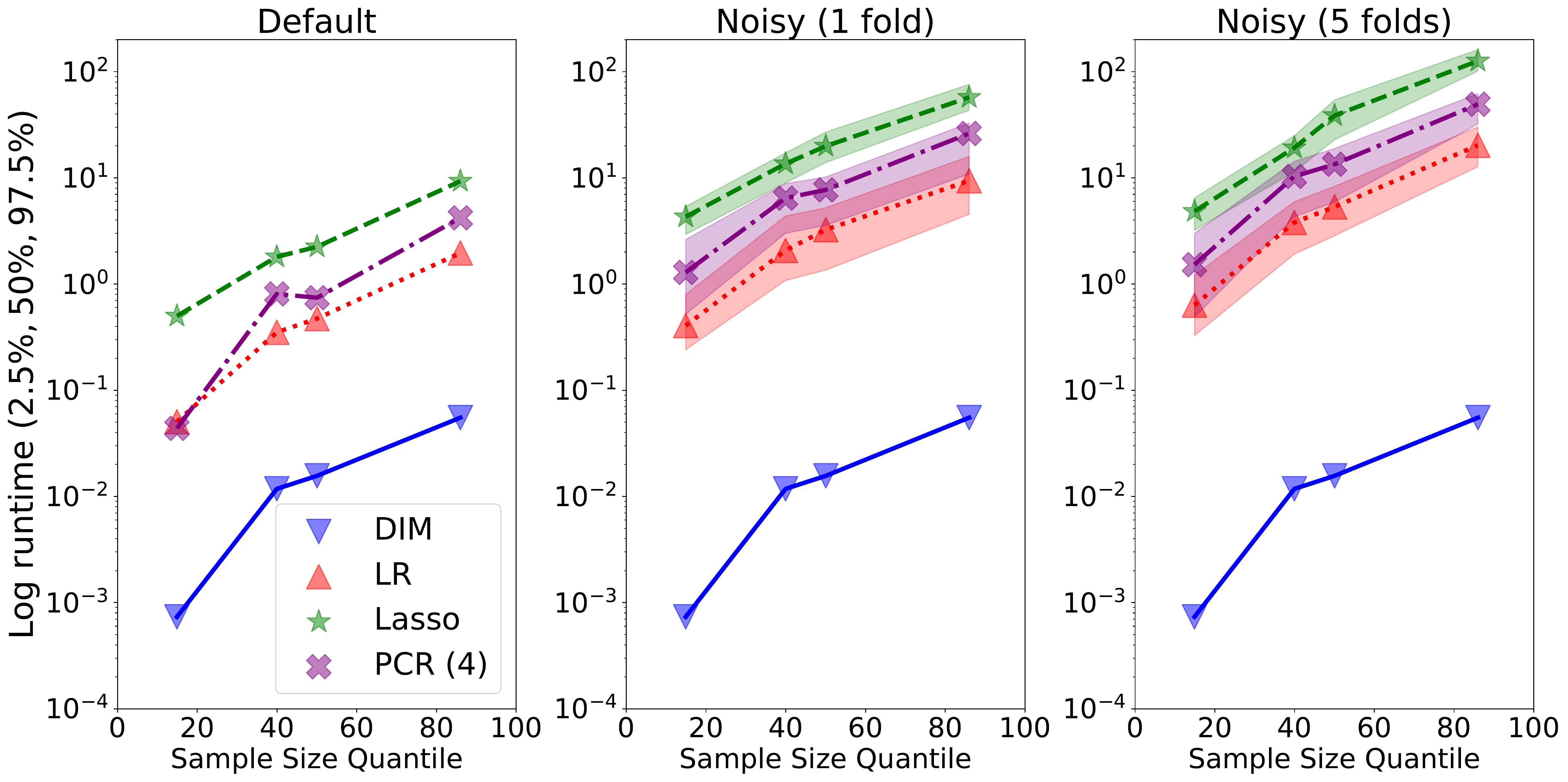}
    \caption{Log computation time (vertical axis) on four representative datasets having sample size $N = F_{N,7}^{-1}(\alpha)$ for $\alpha \in \{0.15, 0.4, .06, 0.85\}$ (horizontal axis). 
    In the left plot we use the default set of $K$ covariates. In the center plot we add one fold of $K$ spurious covariates, in the right plot we add five folds of spurious covariates as per \Cref{eq:spurious}.
    In both these last cases, we repeat the fit $N_{MC}=100$ times on $100$ different randomly drawn noisy sets of covariates.}
    \label{fig:runtime}
\end{figure}

\Cref{eq:spurious} simulates a setting in which we might be using a large set of not curated covariates, some of which are noisy and uncorrelated with the outcomes (violating condition (C2) in \Cref{sec:gobes}). We analyze in \Cref{fig:runtime} the computation cost of running different covariate adjusted methods, relative to the baseline $\DIM$, as a function of the size of the experiment and the number of covariate used. In our experiments, even for larger ones, the computation cost of covariate adjusted methods is moderate. We run experiments using the popular \texttt{scipy} python library \citep{scipy} on a 16-core Intel(R) Xeon(R) CPU E5-2686 v4 @ 2.30GHz. Computation cost increases at faster rate with the sample size than with the covariates' dimensionality. 

\begin{figure}
    \centering
    \includegraphics{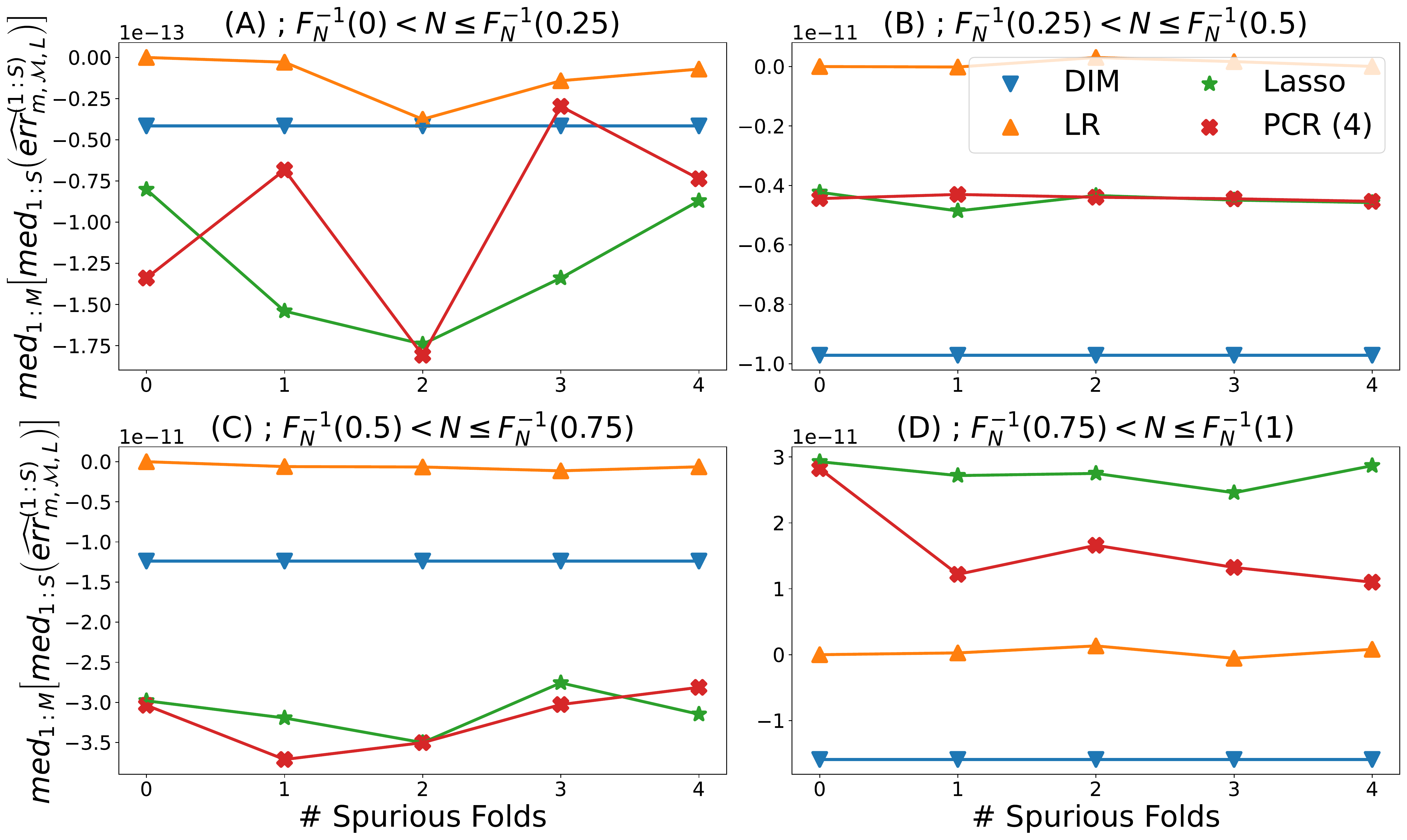}
    \caption{Median of median errors (vertical axis) as a function of the number of spurious folds (horizontal axis) at day $D=7$ of the analysis. Each subplot refers to a different quartile experiments according to their sample size as per $F_{N,7}$. }
    \label{fig:error_distribution}
\end{figure}

\begin{figure}
    \centering
    \includegraphics{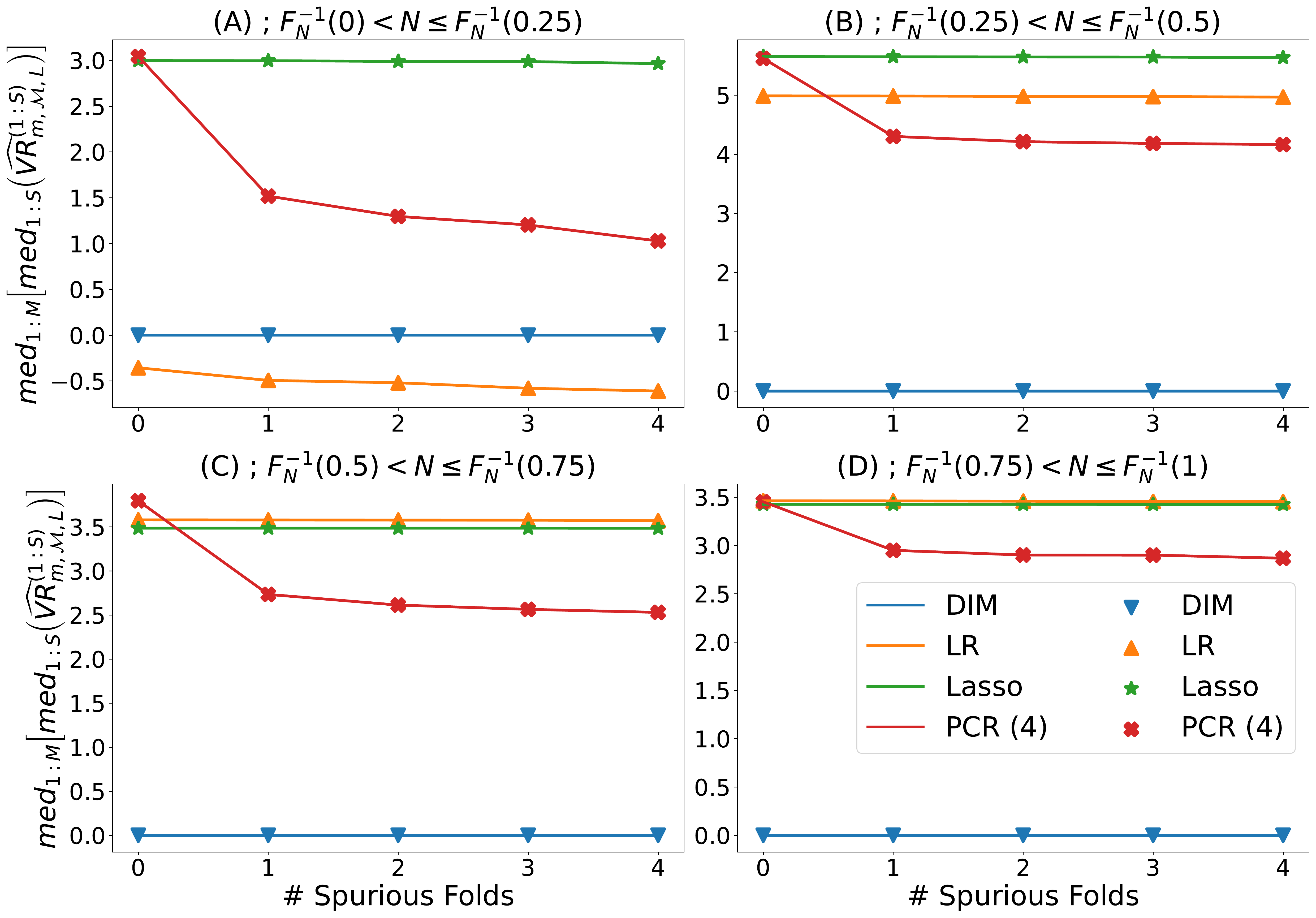}
    \caption{Median of median variance reduction $\hVR_{\MM}$ (vertical axis) as a function of the number of spurious folds (horizontal axis) at day $D=7$ of the analysis. Each subplot refers to a different quartile experiments according to their sample size as per $F_{N,7}$.}
    \label{fig:error_distribution_SE}
\end{figure}

We next assess the robustness of different regression-adjusted methods to the presence of noisy covariates. Because the true causal effect is unknown, we treat the estimate obtained using the linear regression model on the full data as the ``ground truth'' --- i.e., $\ATE:=\hatATE_{\LR}$ with data $\{y_{1:N}, J_{1:N}, \bm{z}_{1:N}\}$. For a fixed $L\ge 1$ and a large number of Monte Carlo random draws $S$, we (i) draw spurious covariates $\tilde{\bm{z}}_{1:N}$ and (ii) compute the empirical distribution of the percentage absolute difference (or error) in the estimate in the presence of spurious covariates with respect to the ground truth:
\[
	\herr_{w,\MM,L}^{(s)}:= \frac{|\hatATE^{(s)}_{w,\MM,L}-\ATE_w|}{|\ATE_w|}.
\]
For experiment $w$, $\hatATE^{(s)}_{w,\MM,L}$ is the estimate of the true $\ATE_w$ using model $\MM$ on the $s$-th random re-draw of covariates with $L$ spurious folds. Results are displayed in \Cref{fig:error_distribution}, where we break down the distribution of the error by clustering experiments in the meta-analysis according to their sample size. Specifically, we divide experiments into four quartiles according to $F_{N,7}$ ((A)--(D)).
     Within each quartile, we compute for each experiment $w$ and for each method $\MM$ and fold $L$ the median error across $s=1,\ldots,S$ ($\text{med}_{1:S}(\herr_{w,\MM,L}^{(1:S)})$). For each quartile of the sample size distribution, this procedure gives us a list of $25$ values for each $\MM, L$. We plot in \Cref{fig:error_distribution} the median across these 25 median errors (vertical axis) as a function of the number of spurious folds (horizontal axis) across different methods.  The estimators considered are extremely robust to noise in terms of point estimates, even in the presence of several noisy covariates. 
Adding spurious noisy covariates has also limited impact on variance reduction gains, as showed in \Cref{fig:error_distribution_SE}. This shows us that the performance of covariate adjusted methods is reliable in the presence of noisy covariates, and the computation cost --- for a well optimized library --- is not prohibitive, even in the presence of large sample sizes.

\section{Discussion} \label{sec:discussion}

In this paper, we have discussed the value and potential of adopting a large class of covariate adjusted models (Generalized Oaxaca-Blinder Estimators) for the estimation of causal effects in online A/B testing. GOBEs rely on a simple but very general procedure, discussed in \Cref{alg:GOBE}. 
By leveraging additional covariates and adopting linear and non-linear regression models, we showed in \Cref{sec:experiments} on extensive experiments on real data that these estimators lead to precise (\Cref{sec:vr}) and robust (\Cref{sec:chance_imbalance}) estimates of the causal effects of interest, which vastly outperform the simple difference in means estimator. Adoption of these estimators can help practitioners understand the effectiveness of the intervention being tested in shorter periods of time, cutting experimentation cost and streamlining the adoption of beneficial innovations (\Cref{sec:duration_recommendation}).
The upfront cost to be paid in order to obtain these more precise estimates is both computational and statistical. We discuss these drawbacks in \Cref{sec:computation}, in which we analyse the performance and computation cost incurred by linear models, and regularized versions thereof on our meta-analysis. In particular, we focus on how such performance scales with the sizes of experiments and dimensionality of covariates. We find that for the models considered, computation cost is not prohibitive even for larger experiments, and inferences are reliable even in the presence of several spurious covariates. 

In light of  the practical concerns and desiderata outlined in \Cref{sec:complexity}, we choose to only consider interpretable and simple linear models, and their regularized versions. We emphasize, however, that the generalized Oaxaca-Blinder framework for the estimation of causal effects introduced in \Cref{sec:gobes} can be straightforwardly applied to complicated, non-linear regression functions (e.g., neural networks). Fitting flexible, nonlinear regression models typically involves solving a complicated, non-convex optimization problem (like the minimization problem of \Cref{eq:minimize}), and might require to employ cross-fitting approaches like the ones discussed in \Cref{sec:complexity} in order to tune regularization parameters.   The design of paradigms to automate these procedure, and related cost-benefit analyses is an active research area. In settings different from the one we considered, flexible non-linear methods have the potential to vastly outperform the simple linear methods here considered. See, e.g., the discussion in \citet{guo2021machine}. 

We envision a number of exciting avenues for future research. On the methodological side, \citet{guo2021generalized} laid the foundations of a framework to provide provable guarantees for a large class of regression models for the estimation of the causal effects. Enlarging the class of models for which these guarantees hold is an exciting avenue for future work \citep{cohen2020no,list2022using}. Additionally, simplifying the conditions necessary for these guarantees to hold, and strengthening their characterization could further increase the popularity of these approaches. On the applied side, the development of pipelines to automate the identification of the optimal regression functions in the presence of large and heterogeneous datasets is a crucial step towards the adoption of these methods at scale. Towards this goal, thorough investigation of the benefits, costs and risks of adopting large, flexible covariate adjusted regression methods are exciting challenges for practitioners in the upcoming years.

\bibliographystyle{ACM-Reference-Format}
\bibliography{references.bib}


\begin{thebibliography}{17}


\ifx \showCODEN    \undefined \def \showCODEN     #1{\unskip}     \fi
\ifx \showDOI      \undefined \def \showDOI       #1{#1}\fi
\ifx \showISBNx    \undefined \def \showISBNx     #1{\unskip}     \fi
\ifx \showISBNxiii \undefined \def \showISBNxiii  #1{\unskip}     \fi
\ifx \showISSN     \undefined \def \showISSN      #1{\unskip}     \fi
\ifx \showLCCN     \undefined \def \showLCCN      #1{\unskip}     \fi
\ifx \shownote     \undefined \def \shownote      #1{#1}          \fi
\ifx \showarticletitle \undefined \def \showarticletitle #1{#1}   \fi
\ifx \showURL      \undefined \def \showURL       {\relax}        \fi
\providecommand\bibfield[2]{#2}
\providecommand\bibinfo[2]{#2}
\providecommand\natexlab[1]{#1}
\providecommand\showeprint[2][]{arXiv:#2}

\bibitem[\protect\citeauthoryear{Camerlenghi, Favaro, Masoero, and
  Broderick}{Camerlenghi et~al\mbox{.}}{2022}]%
        {camerlenghi2022scaled}
\bibfield{author}{\bibinfo{person}{Federico Camerlenghi},
  \bibinfo{person}{Stefano Favaro}, \bibinfo{person}{Lorenzo Masoero}, {and}
  \bibinfo{person}{Tamara Broderick}.} \bibinfo{year}{2022}\natexlab{}.
\newblock \showarticletitle{Scaled process priors for Bayesian nonparametric
  estimation of the unseen genetic variation}.
\newblock \bibinfo{journal}{\emph{J. Amer. Statist. Assoc.}}
  (\bibinfo{year}{2022}), \bibinfo{pages}{1--12}.
\newblock


\bibitem[\protect\citeauthoryear{Cohen and Fogarty}{Cohen and Fogarty}{2020}]%
        {cohen2020no}
\bibfield{author}{\bibinfo{person}{Peter~L Cohen} {and}
  \bibinfo{person}{Colin~B Fogarty}.} \bibinfo{year}{2020}\natexlab{}.
\newblock \showarticletitle{No-harm calibration for generalized
  {O}axaca-{B}linder estimators}.
\newblock \bibinfo{journal}{\emph{arXiv preprint arXiv:2012.09246}}
  (\bibinfo{year}{2020}).
\newblock


\bibitem[\protect\citeauthoryear{Deng, Xu, Kohavi, and Walker}{Deng
  et~al\mbox{.}}{2013}]%
        {deng2013improving}
\bibfield{author}{\bibinfo{person}{Alex Deng}, \bibinfo{person}{Ya Xu},
  \bibinfo{person}{Ron Kohavi}, {and} \bibinfo{person}{Toby Walker}.}
  \bibinfo{year}{2013}\natexlab{}.
\newblock \showarticletitle{Improving the sensitivity of online controlled
  experiments by utilizing pre-experiment data}. In
  \bibinfo{booktitle}{\emph{Proceedings of the sixth ACM international
  conference on Web search and data mining}}. \bibinfo{pages}{123--132}.
\newblock


\bibitem[\protect\citeauthoryear{Guo and Basse}{Guo and Basse}{2021}]%
        {guo2021generalized}
\bibfield{author}{\bibinfo{person}{Kevin Guo} {and} \bibinfo{person}{Guillaume
  Basse}.} \bibinfo{year}{2021}\natexlab{}.
\newblock \showarticletitle{The generalized {O}axaca-{B}linder estimator}.
\newblock \bibinfo{journal}{\emph{J. Amer. Statist. Assoc.}}
  (\bibinfo{year}{2021}).
\newblock


\bibitem[\protect\citeauthoryear{Guo, Coey, Konutgan, Li, Schoener, and
  Goldman}{Guo et~al\mbox{.}}{2021}]%
        {guo2021machine}
\bibfield{author}{\bibinfo{person}{Yongyi Guo}, \bibinfo{person}{Dominic Coey},
  \bibinfo{person}{Mikael Konutgan}, \bibinfo{person}{Wenting Li},
  \bibinfo{person}{Chris Schoener}, {and} \bibinfo{person}{Matt Goldman}.}
  \bibinfo{year}{2021}\natexlab{}.
\newblock \showarticletitle{Machine Learning for Variance Reduction in Online
  Experiments}.
\newblock \bibinfo{journal}{\emph{NeurIPS 2021}} (\bibinfo{year}{2021}).
\newblock


\bibitem[\protect\citeauthoryear{Gupta, Kohavi, Tang, Xu, Andersen, Bakshy,
  Cardin, Chandran, Chen, Coey, et~al\mbox{.}}{Gupta et~al\mbox{.}}{2019}]%
        {gupta2019top}
\bibfield{author}{\bibinfo{person}{Somit Gupta}, \bibinfo{person}{Ronny
  Kohavi}, \bibinfo{person}{Diane Tang}, \bibinfo{person}{Ya Xu},
  \bibinfo{person}{Reid Andersen}, \bibinfo{person}{Eytan Bakshy},
  \bibinfo{person}{Niall Cardin}, \bibinfo{person}{Sumita Chandran},
  \bibinfo{person}{Nanyu Chen}, \bibinfo{person}{Dominic Coey},
  {et~al\mbox{.}}} \bibinfo{year}{2019}\natexlab{}.
\newblock \showarticletitle{Top challenges from the first practical online
  controlled experiments summit}.
\newblock \bibinfo{journal}{\emph{ACM SIGKDD Explorations Newsletter}}
  \bibinfo{volume}{21}, \bibinfo{number}{1} (\bibinfo{year}{2019}),
  \bibinfo{pages}{20--35}.
\newblock


\bibitem[\protect\citeauthoryear{Imbens and Rubin}{Imbens and Rubin}{2015}]%
        {imbens2015causal}
\bibfield{author}{\bibinfo{person}{Guido~W Imbens} {and}
  \bibinfo{person}{Donald~B Rubin}.} \bibinfo{year}{2015}\natexlab{}.
\newblock \bibinfo{booktitle}{\emph{Causal inference in statistics, social, and
  biomedical sciences}}.
\newblock \bibinfo{publisher}{Cambridge University Press}.
\newblock


\bibitem[\protect\citeauthoryear{Jin and Ba}{Jin and Ba}{2021}]%
        {jin2021towards}
\bibfield{author}{\bibinfo{person}{Ying Jin} {and} \bibinfo{person}{Shan Ba}.}
  \bibinfo{year}{2021}\natexlab{}.
\newblock \showarticletitle{Towards Optimal Variance Reduction in Online
  Controlled Experiments}.
\newblock \bibinfo{journal}{\emph{arXiv preprint arXiv:2110.13406}}
  (\bibinfo{year}{2021}).
\newblock


\bibitem[\protect\citeauthoryear{Li and Ding}{Li and Ding}{2017}]%
        {li2017general}
\bibfield{author}{\bibinfo{person}{Xinran Li} {and} \bibinfo{person}{Peng
  Ding}.} \bibinfo{year}{2017}\natexlab{}.
\newblock \showarticletitle{General forms of finite population central limit
  theorems with applications to causal inference}.
\newblock \bibinfo{journal}{\emph{J. Amer. Statist. Assoc.}}
  \bibinfo{volume}{112}, \bibinfo{number}{520} (\bibinfo{year}{2017}),
  \bibinfo{pages}{1759--1769}.
\newblock


\bibitem[\protect\citeauthoryear{Lin}{Lin}{2013}]%
        {lin2013agnostic}
\bibfield{author}{\bibinfo{person}{Winston Lin}.}
  \bibinfo{year}{2013}\natexlab{}.
\newblock \showarticletitle{Agnostic notes on regression adjustments to
  experimental data: Reexamining {F}reedman’s critique}.
\newblock \bibinfo{journal}{\emph{The Annals of Applied Statistics}}
  \bibinfo{volume}{7}, \bibinfo{number}{1} (\bibinfo{year}{2013}),
  \bibinfo{pages}{295--318}.
\newblock


\bibitem[\protect\citeauthoryear{List, Muir, and Sun}{List
  et~al\mbox{.}}{2022}]%
        {list2022using}
\bibfield{author}{\bibinfo{person}{John~A List}, \bibinfo{person}{Ian Muir},
  {and} \bibinfo{person}{Gregory~K Sun}.} \bibinfo{year}{2022}\natexlab{}.
\newblock \bibinfo{booktitle}{\emph{Using Machine Learning for Efficient
  Flexible Regression Adjustment in Economic Experiments}}.
\newblock \bibinfo{type}{{T}echnical {R}eport}. \bibinfo{institution}{National
  Bureau of Economic Research}.
\newblock


\bibitem[\protect\citeauthoryear{Masoero, Camerlenghi, Favaro, and
  Broderick}{Masoero et~al\mbox{.}}{2022}]%
        {masoero2022more}
\bibfield{author}{\bibinfo{person}{Lorenzo Masoero}, \bibinfo{person}{Federico
  Camerlenghi}, \bibinfo{person}{Stefano Favaro}, {and} \bibinfo{person}{Tamara
  Broderick}.} \bibinfo{year}{2022}\natexlab{}.
\newblock \showarticletitle{More for less: predicting and maximizing genomic
  variant discovery via Bayesian nonparametrics}.
\newblock \bibinfo{journal}{\emph{Biometrika}} \bibinfo{volume}{109},
  \bibinfo{number}{1} (\bibinfo{year}{2022}), \bibinfo{pages}{17--32}.
\newblock


\bibitem[\protect\citeauthoryear{Richardson, Liu, McQueen, and
  Hains}{Richardson et~al\mbox{.}}{2022}]%
        {richardson2022bayesian}
\bibfield{author}{\bibinfo{person}{Thomas~S Richardson}, \bibinfo{person}{Yu
  Liu}, \bibinfo{person}{James McQueen}, {and} \bibinfo{person}{Doug Hains}.}
  \bibinfo{year}{2022}\natexlab{}.
\newblock \showarticletitle{A Bayesian Model for Online Activity Sample Sizes}.
  In \bibinfo{booktitle}{\emph{International Conference on Artificial
  Intelligence and Statistics}}. PMLR, \bibinfo{pages}{1775--1785}.
\newblock


\bibitem[\protect\citeauthoryear{Rubin}{Rubin}{1977}]%
        {rubin1977assignment}
\bibfield{author}{\bibinfo{person}{Donald~B Rubin}.}
  \bibinfo{year}{1977}\natexlab{}.
\newblock \showarticletitle{Assignment to treatment group on the basis of a
  covariate}.
\newblock \bibinfo{journal}{\emph{Journal of educational Statistics}}
  \bibinfo{volume}{2}, \bibinfo{number}{1} (\bibinfo{year}{1977}),
  \bibinfo{pages}{1--26}.
\newblock


\bibitem[\protect\citeauthoryear{Splawa-Neyman, Dabrowska, and
  Speed}{Splawa-Neyman et~al\mbox{.}}{1990}]%
        {splawa1990application}
\bibfield{author}{\bibinfo{person}{Jerzy Splawa-Neyman},
  \bibinfo{person}{Dorota~M Dabrowska}, {and} \bibinfo{person}{TP Speed}.}
  \bibinfo{year}{1923/1990}\natexlab{}.
\newblock \showarticletitle{On the application of probability theory to
  agricultural experiments. Essay on principles. Section 9.}
\newblock \bibinfo{journal}{\emph{Statist. Sci.}} (\bibinfo{year}{1923/1990}),
  \bibinfo{pages}{465--472}.
\newblock


\bibitem[\protect\citeauthoryear{Tukey}{Tukey}{1991}]%
        {tukey1991use}
\bibfield{author}{\bibinfo{person}{John~W Tukey}.}
  \bibinfo{year}{1991}\natexlab{}.
\newblock \showarticletitle{Use of many covariates in clinical trials}.
\newblock \bibinfo{journal}{\emph{International Statistical Review/Revue
  Internationale de Statistique}} (\bibinfo{year}{1991}),
  \bibinfo{pages}{123--137}.
\newblock


\bibitem[\protect\citeauthoryear{Virtanen, Gommers, Oliphant, Haberland, Reddy,
  Cournapeau, Burovski, Peterson, Weckesser, Bright, {van der Walt}, Brett,
  Wilson, Millman, Mayorov, Nelson, Jones, Kern, Larson, Carey, Polat, Feng,
  Moore, {VanderPlas}, Laxalde, Perktold, Cimrman, Henriksen, Quintero, Harris,
  Archibald, Ribeiro, Pedregosa, {van Mulbregt}, and {SciPy 1.0
  Contributors}}{Virtanen et~al\mbox{.}}{2020}]%
        {scipy}
\bibfield{author}{\bibinfo{person}{Pauli Virtanen}, \bibinfo{person}{Ralf
  Gommers}, \bibinfo{person}{Travis~E. Oliphant}, \bibinfo{person}{Matt
  Haberland}, \bibinfo{person}{Tyler Reddy}, \bibinfo{person}{David
  Cournapeau}, \bibinfo{person}{Evgeni Burovski}, \bibinfo{person}{Pearu
  Peterson}, \bibinfo{person}{Warren Weckesser}, \bibinfo{person}{Jonathan
  Bright}, \bibinfo{person}{St{\'e}fan~J. {van der Walt}},
  \bibinfo{person}{Matthew Brett}, \bibinfo{person}{Joshua Wilson},
  \bibinfo{person}{K.~Jarrod Millman}, \bibinfo{person}{Nikolay Mayorov},
  \bibinfo{person}{Andrew R.~J. Nelson}, \bibinfo{person}{Eric Jones},
  \bibinfo{person}{Robert Kern}, \bibinfo{person}{Eric Larson},
  \bibinfo{person}{C~J Carey}, \bibinfo{person}{{\.I}lhan Polat},
  \bibinfo{person}{Yu Feng}, \bibinfo{person}{Eric~W. Moore},
  \bibinfo{person}{Jake {VanderPlas}}, \bibinfo{person}{Denis Laxalde},
  \bibinfo{person}{Josef Perktold}, \bibinfo{person}{Robert Cimrman},
  \bibinfo{person}{Ian Henriksen}, \bibinfo{person}{E.~A. Quintero},
  \bibinfo{person}{Charles~R. Harris}, \bibinfo{person}{Anne~M. Archibald},
  \bibinfo{person}{Ant{\^o}nio~H. Ribeiro}, \bibinfo{person}{Fabian Pedregosa},
  \bibinfo{person}{Paul {van Mulbregt}}, {and} \bibinfo{person}{{SciPy 1.0
  Contributors}}.} \bibinfo{year}{2020}\natexlab{}.
\newblock \showarticletitle{{{SciPy} 1.0: Fundamental Algorithms for Scientific
  Computing in Python}}.
\newblock \bibinfo{journal}{\emph{Nature Methods}}  \bibinfo{volume}{17}
  (\bibinfo{year}{2020}), \bibinfo{pages}{261--272}.
\newblock
\urldef\tempurl%
\url{https://doi.org/10.1038/s41592-019-0686-2}
\showDOI{\tempurl}


\end{thebibliography}
\end{document}